%% file: main.tex
\documentclass[10pt, aps, prx, twocolumn, superscriptaddress, nofootinbib, noeprint]{revtex4-2}

\usepackage{amsmath, amssymb, bm, mathtools, physics, array, lipsum, graphicx, tikz, placeins}

\usepackage{bbm}
\usepackage[hidelinks]{hyperref}
\usepackage{cleveref}
\usetikzlibrary{quantikz2, arrows}

\usepackage{nicematrix}
\NiceMatrixOptions{small}
\usepackage{booktabs}

\usepackage{amsthm}
\let\vec\bm

\newtheoremstyle{myRemark}{}{}{}{}{\bfseries}{.}{ }{\thmname{#1}\thmnumber{ #2}\normalfont\itshape\thmnote{ (#3)}}

\theoremstyle{myRemark} 

\usepackage{subcaption}
\newcolumntype{L}{>{$}l<{$}}
\newcolumntype{C}{>{$}c<{$}}
\newcolumntype{R}{>{$}r<{$}}

\newcommand{\appropto}{\mathrel{\vcenter{
  \offinterlineskip\halign{\hfil$##$\cr
    \propto\cr\noalign{\kern2pt}\sim\cr\noalign{\kern-2pt}}}}}

\begin{document}

\title{General Photon Subtraction from a Gaussian Perspective}

\author{Niklas Budinger}
\email{nbudinge@t-online.de}
\affiliation{Johannes-Gutenberg University of Mainz, Institute of Physics, Staudingerweg 7, 55128 Mainz, Germany}
\affiliation{Center for Macroscopic Quantum States (bigQ), Department of Physics, Technical University of Denmark, 2800 Kongens Lyngby, Denmark}

\author{Ulrik L. Andersen}
\affiliation{Center for Macroscopic Quantum States (bigQ), Department of Physics, Technical University of Denmark, 2800 Kongens Lyngby, Denmark}

\author{Peter~van~Loock}
\email{loock@uni-mainz.de}
\affiliation{Johannes-Gutenberg University of Mainz, Institute of Physics, Staudingerweg 7, 55128 Mainz, Germany}

\date{\today}

\begin{abstract}
A near-deterministic source of non-Gaussianity is the missing piece to reach universal and fault-tolerant quantum computing with continuous-variable optics.
Due to a lack of strong non-linearities, the probabilistic photon-subtracted Gaussian states remain the most promising contender. In combination with Gaussian measurements and single-mode feed-forward operations, they have been shown to provide sufficient non-Gaussianity at high rates. However, these existing breeding protocols lack the necessary loss tolerance to be experimentally feasible.
In this work, we introduce a mathematical formalism based on the application of a Gaussian blur and filter on the measured Fock Wigner function to describe general photon-subtracted Gaussian states. Within our representation, the initial squeezing, Gaussian measurements, and photon loss can all be understood as contributions to the total blurring and interchanged accordingly.
We find that this intuitive approach can be used to find improvements to cat, cubic phase, and GKP state generation by compromising between success rates, feed-forward compatibility, and loss tolerance. In the case of multiple photon subtractions, this is achieved by introducing a \textit{hybrid} setup variant bridging the gap between protocols based on breeding and post-selection. Finally, we establish a general lower bound on the resources needed to generate a given target state near-deterministically by introducing the expected Wigner logarithmic negativity as an additive monotone of non-Gaussianity.

\end{abstract}

\maketitle

\input{introduction}
\input{state_representation}
\input{applications}
\input{conclusion}

\acknowledgments
We gratefully acknowledge support from the Danish National Research Foundation (bigQ, DNRF0142), EU project CLUSTEC (grant agreement no. 101080173), EU ERC project ClusterQ (grant agreement no. 101055224, ERC2021-ADG), Innovation Fund Denmark (QuantERA - ClusSTAR, 3155-00024A), and funding from the BMFTR in Germany (QR.N, QuKuK, QuaPhySI, PhotonQ).

\bibliography{references}

\onecolumngrid
\appendix
\input{appendix}

\end{document}

%% file: introduction.tex
\section{Introduction}

Continuous-variable quantum optics is a serious contender to achieve universal and fault-tolerant quantum computing via large-scale measurement-based cluster states \cite{zhangContinuousvariableGaussianAnalog2006,menicucciUniversalQuantumComputation2006,tzitrinFaulttolerantQuantumComputation2021,baragiolaAllGaussianUniversalityFault2019a,ostergaardOctoRailLatticeFourdimensional2025}. Respective architectures have been realised both on chip \cite{aghaeeradScalingNetworkingModular2025} and in tabletop experiments \cite{yokoyamaUltraLargeScaleContinuousVariableCluster2013,yoshikawaGenerationOnemillionmodeContinuousvariable2016,larsenDeterministicGenerationTwodimensional2019,yokoyamaFullstackAnalogOptical2026} leveraging the low decoherence, room temperature operation and temporal multiplexing provided by linear optical components.
In combination with photon-number resolving detectors (PNRDs), these have led to demonstrations of
near-term quantum advantage in the form of Gaussian Boson Sampling (GBS) \cite{liuGaussianBosonSampling2026,madsenQuantumComputationalAdvantage2022a,hamiltonGaussianBosonSampling2017,aaronsonComputationalComplexityLinear2010} with potential applications ranging from identifying dense subgraphs \cite{bromleyApplicationsNearTermPhotonic2020} to solving Gaussian expectation problems \cite{andersenUsingGaussianBoson2025a,andersenEstimatingPercentageGBS2025a}.

On the other hand, the near-deterministic generation of non-Gaussian resource states needed for universality and fault-tolerance remains elusive.
Due to a lack of strong optical non-linearities, 
the most promising approaches are GBS-like setups relying on Gaussian states and PNRDs \cite{suGenerationPhotonicNonGaussian2019,suConversionGaussianStates2019a}.
In order to handle their inherent complexity, optimisation techniques based on the loop hafnian \cite{quesadaSimulatingRealisticNonGaussian2019,miattoFastOptimizationParametrized2020,yaoRiemannianOptimizationPhotonic2024}, coherent state decomposition \cite{marshallSimulationQuantumOptics2023a,solodovnikovaFastSimulationsContinuousvariable2025a}, backcasting \cite{fukuiEfficientBackcastingSearch2022a}, as well as machine-learning \cite{sabapathyProductionPhotonicUniversal2019,odriscollHybridMachineLearning2019,tzitrinProgressPracticalQubit2020a} have been developed.
However, the resulting setups depend on the measurement of specific photon patterns and are probabilistic in nature.
In contrast, the near-deterministic generation of cat \cite{ourjoumtsevGenerationOpticalSchrodinger2007,takaseGenerationOpticalSchrodinger2021} and cubic phase states \cite{gottesmanEncodingQubitOscillator2001,ghoseNonGaussianStatesContinuous2006} has been proposed using high photon numbers along with Gaussian feed-forward operations.
Protocols based on breeding \cite{weigandGeneratingGridStates2018,takaseGenerationFlyingLogical2024} and gate teleportation \cite{gottesmanEncodingQubitOscillator2001,sakaguchiNonlinearFeedforwardEnabling2022,budingerAllopticalQuantumComputing2024a} can then be used to produce certain more complex non-Gaussian states -- such as GKP \cite{gottesmanEncodingQubitOscillator2001} states -- at high rates. Yet, these have been found to quickly diminish under realistic levels of photon loss \cite{aghaeeradScalingNetworkingModular2025,solodovnikovaLossToleranceCat2025}.

In order to overcome these limitations, theoretical improvements driven by a better understanding of the interplay of success rate, feed-forwards, and photon loss are necessary, alongside further experimental advances \cite{larsenIntegratedPhotonicSource2025b,yuExtensibleUniversalPhotonic2026,endoPicosecondSchrodingerCat2026}.
In this context, recent results include the addition of displacements to cat state generation \cite{notarnicolaDeterministicFeedforwardbasedGeneration2026},
the near-deterministic machine-learning-based preparation of cat and cubic phase states \cite{antenehMachineLearningEfficient2024,antenehDeepReinforcementLearning2026},
the stellar decomposition \cite{motamediStellarDecompositionGaussian2026},
as well as the distillation of non-Gaussian control parameters \cite{hanamuraStellarRankControl2025}.

In this work, we introduce a novel mathematical formalism designed to describe general GBS-like setups and their generated output states.
Based on the application of a Gaussian blur and a Gaussian filter on the Wigner function of the measured Fock states, it provides an intuitive understanding of the produced Wigner negativity and thus non-Gaussianity of a setup.
Along with the expected Wigner logarithmic negativity as a natural figure of merit, we use it to analyse the relation of the average created non-Gaussianity and feed-forward compatibility for general GBS-like setups under different types of photon loss.

In the case of one PNRD, we find that the relevant state space can be described by only two parameters and is thus fully assessable.
We use this to introduce a general lower bound on the resources needed to generate a given target state near-deterministically, as well as two distinct single-mode feed-forward schemes for cat- and cubic phase-like states. Furthermore, we discuss setup optimisations which compromise between non-Gaussianity, feed-forward compatibility, and loss tolerance.
In the case of multiple PNRDs, we introduce a \textit{hybrid} setup variant which combines features of the post-selection and breeding protocols and allows for general feed-forward operations.
In addition, we show how and when it can be used to continuously morph one protocol into the other, thereby compromising between individual success rates and feed-forward compatibility.
Finally, we test the performance of the \textit{hybrid} setup variant for GKP state generation and highlight how
the concept of a Gaussian blur generated by the initial squeezing, Gaussian measurements and photon loss
can be utilised to adjust the success rate, feed-forward compatibility, and loss tolerance of a general GBS-like setup to one's liking.

This paper is structured as follows: Section~\ref{sec:StateRepresentation} introduces the mathematical formalism describing the photon-subtracted Gaussian states as well as general notation. The expected Wigner logarithmic negativity is then defined in Sec.~\ref{sec:xWLN}. Section~\ref{sec:MaximumxWLN} explores the relevant state space of setups with one PNRD,
while the resource bound and single-mode feed-forwards are covered in Sec.~\ref{sec:SingleModeFF}.
Finally, setups with multiple PNRDs are discussed in Sec.~\ref{sec:GeneralMultiModeFF}.

%% file: state_representation.tex
\section{State Representation}\label{sec:StateRepresentation}
First, let us go through the different relevant states and their notation used throughout this paper.
To avoid ambiguity, we refrain from using the term `general photon subtraction' which was prominently used in a more narrow sense in \cite{takaseGenerationOpticalSchrodinger2021}. Instead, we refer to the states produced by measuring a Gaussian resource with PNRDs as photon-subtracted Gaussian (PSG) states and to the corresponding experimental setups as GBS-like.
Besides, we use the convention $\hbar= 2$.

\subsection{General Gaussian state}\label{sec:GeneralGaussianState}
The Wigner function of a general $(k+m)$-mode Gaussian state is given by
\begin{align}
    W_{\Sigma, \vec{\mu}}(\vec{q})=\left(\det 2\pi\Sigma\right)^{-\frac{1}{2}}\cdot G_{\Sigma, \vec{\mu}}(\vec{q})
\end{align}
with the unnormalised Gaussian function
\begin{align}
    G_{\Sigma, \vec{\mu}}(\vec{q})=\exp\left(-\frac{1}{2}(\vec{q}-\vec{\mu})^T\Sigma^{-1}(\vec{q}-\vec{\mu})\right),
\end{align}
the vector of quadratures $\vec{q} = \left(x_1, p_1, ..., x_{k+m}, p_{k+m}\right)^T$ and the displacement $\vec{\mu}\in\mathbb{R}^{2(k+m)}$. The covariance matrix $\Sigma\in\mathbb{R}^{2(k+m)\times 2(k+m)}$ is required to be symmetric $\Sigma^T=\Sigma$, positive definite $\Sigma>0$ and fulfil the uncertainty relation
\begin{align}
    \Sigma + i \Omega_{k+m}\geq0
\end{align}
where $\Omega_{k+m}=I_{k+m}\otimes \left[\begin{smallmatrix}0&1\\-1&0\end{smallmatrix}\right]$ is the symplectic form and  $I$ is the identity matrix.
We will label the $k$- and the $m$-mode subsystem by $A$ and $B$, respectively, and denote the partition of covariance matrix and displacement as
\begin{flalign}\label{eq:Partition}
    &&\Sigma=\begin{bmatrix}\Sigma_A&\Sigma_{AB}\\\Sigma_{AB}^T&\Sigma_B\end{bmatrix}&&\text{and}&&
    \vec{\mu}=\begin{bmatrix}\vec{\mu}_A\\\vec{\mu}_B\end{bmatrix}.
    &&\begin{array}{l}
        \left.\vphantom{\vec{\mu}_A}\right]2k\\
        \left.\vphantom{\vec{\mu}_B}\right]2m
    \end{array}
\end{flalign}
The resulting Gaussian state is pure iff $\det(\Sigma)=1$.

\subsection{General Fock state}\label{sec:GeneralFockState}
The Wigner function of a single-mode Fock state with $n$ photons is given by
\begin{align}
    W_{n}(x, p)=\frac{(-1)^{n}}{2\pi} \exp\left({-\frac{x^2+p^2}{2}}\right)\cdot L_{n}(x^2+p^2)
\end{align}
with the Laguerre polynomial
\begin{align}
    L_{n}(y)=\sum_{j=0}^{n} \binom{n}{j}\frac{(-1)^j}{j!} y^j.
\end{align}
A $k$-mode Fock state with photon numbers $\vec{n}\in\mathbb{N}^k_0$ will be denoted by
\begin{align}
    W_{\vec{n}}(\vec{q})=\prod_{j=1}^kW_{n_j}(x_j, p_j).
\end{align}

\subsection{Photon-subtracted Gaussian state}\label{sec:PhotonSubtractedGaussianState}

\begin{figure*}[t]
    \centering
    \includegraphics[width=1\textwidth]{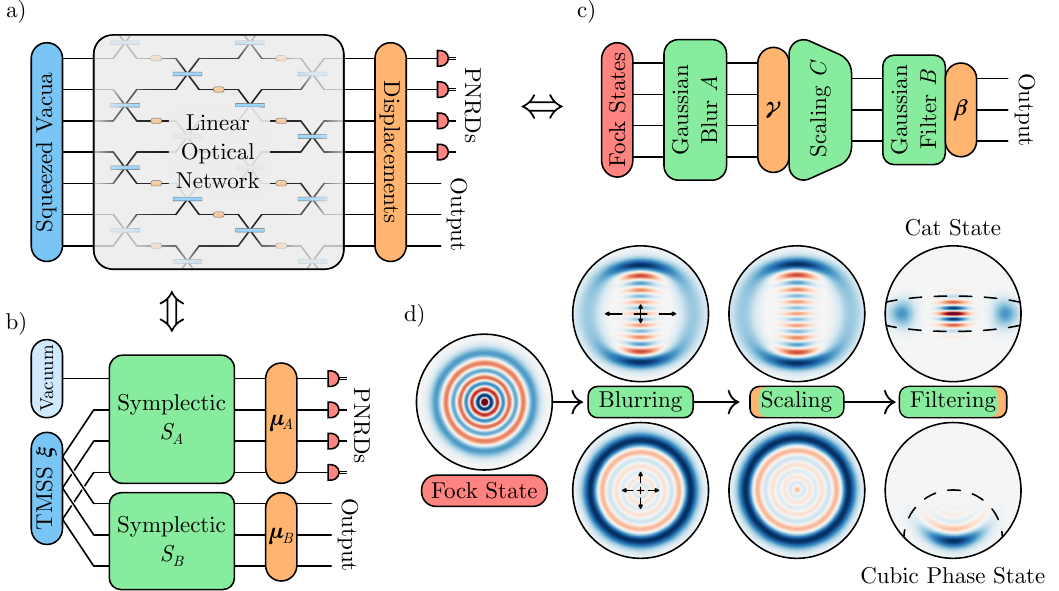}
    \caption{Schematic representation of the GBS-like experimental setup as well as equivalent theoretical descriptions used for generating a pure photon-subtracted Gaussian (PSG) state. a) In the experiment, squeezed vacua are sent through a passive linear optical network followed by potential displacements. A subset of modes is measured by photon-number resolving detectors (PNRDs) resulting in a PSG state dependent on the measured photon number pattern. b) An array of two-mode squeezed states (TMSS) is split into two paths which are filled up with single-mode vacuum if necessary. The top path undergoes the symplectic operation $S_A$ and the displacement $\mu_A$ before being measured by the PNRDs. The output state is then obtained by applying the symplectic operation $S_B$ as well as the displacement $\mu_B$ on the bottom path. c) The Wigner function of the measured Fock states is subjected to the Gaussian blur $A$, the coordinate transform $(C, \vec{\gamma})$, and the Gaussian filter and final displacement $(B, \vec{\beta})$ producing the Wigner function of the PSG state. d) Depiction of the specific blurring, scaling, and filtering required to produce a cat-like as well as a cubic phase-like state and their effect on the Fock state $\ket{n=7}$. The size of the diagonal entries of $A$ ($B$) is indicated by the arrows (ellipses), while setting $S_A=S_B$ minimizes the effect of $C\propto Z_2$. The generalisation to mixed states is straightforward for all descriptions but b).}
    \label{fig:Schemata}
\end{figure*}

Detecting this photon pattern $\vec{n}$ on the first $k$ modes of a $(k+m)$-mode Gaussian state produces the Wigner function of the PSG state
\begin{align}\label{eq:GaussianPSG_0}
    W_{\Sigma, \vec{\mu}, \vec{n}}(\vec{q})= \frac{(4\pi)^k}{P_{\vec{n}}\sqrt{\det 2\pi\Sigma}}\int_{\mathbb{R}^{2k}} d\vec{r}\ G_{\Sigma, \vec{\mu}}\left(\begin{bmatrix}\vec{r}\\ \vec{q}\end{bmatrix}\right)W_{\vec{n}}(\vec{r})
\end{align}
where $P_{\vec{n}}$ is the probability of measuring said photon pattern.
Typically, this Wigner function is then approximated by either writing the Gaussian state $G_{\Sigma, \vec{\mu}}$ in Fock basis and choosing an appropriate cut-off \cite{quesadaSimulatingRealisticNonGaussian2019,suGenerationPhotonicNonGaussian2019,suConversionGaussianStates2019a}, or by representing the Fock Wigner function $W_{\vec{n}}$ as a sum of Gaussians \cite{marshallSimulationQuantumOptics2023a,solodovnikovaFastSimulationsContinuousvariable2025a}.
In this work, we propose an alternative approach:
Using the partition from Eq.~\eqref{eq:Partition}, we can represent $W_{\Sigma, \vec{\mu}, \vec{n}}$ by a Gaussian convolution of the Wigner function $W_{\vec{n}}$, followed by a Gaussian filter
\begin{align}\label{eq:GaussianPSG}
    W_{\Sigma, \vec{\mu}, \vec{n}}(\vec{q})&=\frac{N_\Sigma}{P_{\vec{n}}}\cdot G_{B, \vec{\beta}}\left(\vec{q}\right)\cdot \big(G_{A, \vec{0}}\ast W_{\vec{n}}\big)\left(C\vec{q}+\vec{\gamma}\right).
\end{align}
Here, the Gaussian which the initial $k$-mode Fock state is convoluted with is characterised by the Schur complement
\begin{align}
    A &= \Sigma/\Sigma_B := \Sigma_A - \Sigma_{AB}\Sigma_B^{-1}\Sigma_{AB}^T,
\end{align}
the Gaussian filter is given by the covariance matrix and mean of subsystem $B$
\begin{align}
    B = \Sigma_B,&&
    \vec{\beta} = \vec{\mu}_B,
\end{align}
and the coordinate transform in between is dependent on the cross-term
\begin{align}
    C = \Sigma_{AB}\Sigma_B^{-1},&&
    \vec{\gamma} = \vec{\mu}_A - \Sigma_{AB}\Sigma_B^{-1}\vec{\mu}_B.
\end{align}
Finally, the normalisation constant is given by
\begin{align}
    N_\Sigma &= \frac{2^k(2\pi)^{-m}}{\sqrt{\det\Sigma}}=\frac{2^k(2\pi)^{-m}}{\sqrt{\det A \det B}},
\end{align}
ensuring that both the Gaussian convolution and filter are normalised.
The derivation of Eq.~\eqref{eq:GaussianPSG} can be found in App.~\ref{sec:AppendixDerivation}. Figure~\ref{fig:Schemata}c presents a schematic depiction of this PSG state representation, while Fig.~\ref{fig:Schemata}d highlights the specific blurring, scaling, and filtering needed to produce a cat-like as well as a cubic phase-like state.

\paragraph*{Pure states.}\label{sec:PureStates}
In case the initial Gaussian state is pure, we can make use of the canonical form found in \cite{boteroModeWiseEntanglementGaussian2003,giedkeEntanglementTransformationsPure2003} in order to simplify the three matrices $A$, $B$, and $C$.
Therefore, the covariance matrix $\Sigma$ is written in terms of the covariance matrix of an array of $\ell=\rank\left(\Sigma_{AB}\right)$ two-mode squeezed states filled up with additional vacuum modes,
\begin{flalign}
    &&\Sigma_\text{TMSS}=\left[
    \begin{smallmatrix}
        C_h\otimes I_2 &  & S_h\otimes Z_2 &  \\
         & I_{2k-2\ell} &  & 0 \\
        S_h\otimes Z_2 &  & C_h\otimes I_2 &  \\
         & 0 &  & I_{2m-2\ell}
    \end{smallmatrix}
    \right],&&
    \begin{array}{l}
        \left.\vphantom{\begin{smallmatrix}I_2\\I_2\end{smallmatrix}}\right]2k\\[1mm]
        \left.\vphantom{\begin{smallmatrix}I_2\\I_2\end{smallmatrix}}\right]2m
    \end{array}
\end{flalign}
as well as two symplectic matrices $S_A\in\text{Sp}(2k, \mathbb{R})$ and $S_B\in\text{Sp}(2m, \mathbb{R})$ which act separately on the two subsystems,
\begin{flalign}\label{eq:CanonicalForm}
    &&\Sigma=\begin{bmatrix}S_A&\\&S_B\end{bmatrix}\Sigma_\text{TMSS}\begin{bmatrix}S_A^T&\\&S_B^T\end{bmatrix}.&&
    \begin{array}{l}
        \left.\vphantom{S_A}\right]2k\\[0.3mm]
        \left.\vphantom{S_A}\right]2m
    \end{array}
\end{flalign}
Here, we used the notation $Z_2=\left[\begin{smallmatrix}1&0\\0&-1\end{smallmatrix}\right]$ and
\begin{align}
    C_h=\left[\begin{smallmatrix}\cosh \xi_1&&\\&\ddots&\\&&\cosh \xi_\ell\end{smallmatrix}\right],\ 
    S_h=\left[\begin{smallmatrix}\sinh \xi_1&&\\&\ddots&\\&&\sinh \xi_\ell\end{smallmatrix}\right]
\end{align}
where $\vec{\xi}\in\left(\mathbb{R}\setminus \{0\}\right)^\ell$ provides the amount of two-mode squeezing.
The parameters to Eq.~\eqref{eq:GaussianPSG} then become
\begin{align}
    A &= S_A\left(\left(C_h^{-1}\otimes I_2\right)\oplus I_{2k-2\ell}\right)S_A^T,\nonumber\\
    B &= S_B\left(\left(C_h\otimes I_2\right)\oplus I_{2m-2\ell}\right)S_B^T,\\
    C &= S_A\left(\left(C_h^{-1}S_h\otimes Z_2\right)\oplus 0_{(2k-2\ell)\times(2m-2\ell)}\right)S_B^{-1},\nonumber
\end{align}
as well as
\begin{align}
    \vec{\beta} &= \vec{\mu}_B,&
    \vec{\gamma} &= \vec{\mu}_A - C\vec{\mu}_B,&
    N_\Sigma &= 2^k(2\pi)^{-m}.
\end{align}
This description of a pure PSG state is depicted in Fig.~\ref{fig:Schemata}b.

\paragraph*{Photon loss.}\label{sec:PhotonLoss}
In experimental realisations, the initial Gaussian is implemented by generating offline squeezed states that are interfered at a passive linear optical network and subsequently displaced before being measured by photon-number resolving detectors (PNRDs), see Fig.~\ref{fig:Schemata}a.
The most prominent source of error in this GBS-like setup is photon loss which we will consider in three specific locations: i) on the offline squeezed states, ii) before the photon detectors, and iii) on the output state.
Since uniform loss acting on the offline squeezed states commutes with the passive linear optical network and trivially scales the displacements, it can be combined with the detector and output loss, respectively. Then, the accumulated loss acts as a single Gaussian channel on the initially pure Gaussian state
\begin{align}\begin{split}
    \Sigma_\text{loss}&=\sqrt{H}\Sigma_\text{pure}\sqrt{H}+(I_{k+m}-H)\\
    \vec{\mu}_\text{loss}&=\sqrt{H}\vec{\mu}_\text{pure}
\end{split}\end{align}
where
\begin{align}
    H=
    \begin{bmatrix}
        \eta_A I_k & \\
         & \eta_B I_m 
    \end{bmatrix}.
\end{align}
The two parameters $0\leq\eta_A, \eta_B\leq1$ represent the sum of offline/detector and offline/output loss, respectively.
The parameters to Eq.~\eqref{eq:GaussianPSG} then become
\begin{align}\begin{split}
    A_\text{loss} ={}& \eta_A A_\text{pure} + (1-\eta_A)I_k + \eta_A(\eta_B^{-1}-1)\\
    &\times C_\text{pure}\left(I_m+(\eta_B^{-1}-1)B_\text{pure}^{-1}\right)^{-1}C_\text{pure}^T,\\
    B_\text{loss} ={}& \eta_B B_\text{pure} + (1-\eta_B)I_m,\\
    C_\text{loss} ={}& \sqrt{\frac{\eta_A}{\eta_B}}C_\text{pure}\left(I_m+(\eta_B^{-1}-1)B_\text{pure}^{-1}\right)^{-1}.
\end{split}\end{align}
While corresponding formulas for $\vec{\beta}_\text{loss}$ and $\vec{\gamma}_\text{loss}$ can be obtained, displacements can easily be adjusted to account for loss. Thus in the interest of simplicity, we will instead assume $\vec{\mu}_\text{loss}=\vec{\mu}_\text{pure}$ and consequently use
\begin{align}
    \vec{\beta}_\text{loss} = \vec{\mu}_B&&\text{and}&&
    \vec{\gamma}_\text{loss} = \vec{\mu}_A - C_\text{loss}\vec{\mu}_B
\end{align}
from here on.
In the common case that $B_\text{pure}\gg1$ (i.e.\ its eigenvalues are much greater than one) the effect of loss can be simplified to
\begin{align}
    B_\text{loss} \approx{}& \eta_B B_\text{pure},&
    C_\text{loss} \approx{}& \sqrt{\frac{\eta_A}{\eta_B}}C_\text{pure},
\end{align}
as well as
\begin{align}
    A_\text{loss} \approx{}& \eta_A A_\text{pure} + (1-\eta_A)I_k + \eta_A(\eta_B^{-1}-1)C_\text{pure}C_\text{pure}^T\nonumber\\
    ={}&\eta_A S_A\Big[\left(C_h^{-1}\otimes I_2\right)\oplus I_{2k-2\ell} + \left(\eta_A^{-1}-1\right)S_A^{-1}S_A^{-T}\nonumber\\
    &\qquad\quad + \left(\eta_B^{-1}-1\right)Z_0S_B^{-1}S_B^{-T}Z_0^T\Big]S_A^T,
    \label{eq:LossApproximation}
\end{align}
where $(\cdot)^{-T}$ denotes the inverse transpose and $Z_0\equiv\left(I_\ell\otimes Z_2\right)\oplus 0_{(2k-2\ell)\times(2m-2\ell)}$.
This allows for a direct comparison of the different loss sources: While the size of the Wigner function in phase space is scaled by a factor of $\sqrt{\eta_B/\eta_A}$, the additional loss-induced blur is generally higher for $\eta_B$ whenever the two symplectic matrices $S_A$ and $S_B$ align. However, given that any choice of $S_B$ can be undone by a Gaussian operation later on, it can be purposefully chosen to minimize the impact of $\eta_B$ on the output, rendering it less consequential than $\eta_A$ in certain cases. In the following, both loss parameters are chosen to be equal $\eta_A=\eta_B=\eta$ unless otherwise stated.
The implementation of other Gaussian noise channels is straightforward within the presented formalism, but not further explored within this work.

%% file: applications.tex
\section{Applications}\label{sec:Applications}
Generating any non-Gaussian features using photon measurements inherently comes with a success probability $P_{\vec{n}}<1$. Moreover, directly obtaining a specific non-Gaussian state is generally unlikely. Near-deterministic state generation therefore requires setups that produce similar outputs with high probability. A combination of post-selection and Gaussian feed-forwards can then vastly increase the overall success rate, as demonstrated for cat-like \cite{ourjoumtsevGenerationOpticalSchrodinger2007,takaseGenerationFlyingLogical2024} and cubic phase-like states \cite{gottesmanEncodingQubitOscillator2001,ghoseNonGaussianStatesContinuous2006}.
Besides, spatial (or temporal) multiplexing can be used to further increase probabilities at the cost of a linear scaling of resources \cite{aghaeeradScalingNetworkingModular2025,takaseGenerationFlyingLogical2024}.
However, high overall success rates still necessitate high photon numbers, which increase a setup's susceptibility to photon loss and in turn reduce its outputs' non-Gaussianity.

This complex interplay of non-Gaussianity, success probability, and loss tolerance makes it difficult to fine-tune setups intended to produce specific non-Gaussian states. In the following, we explore their interdependency using the presented description of PSG state generation. Hereto, we first define the expected Wigner logarithmic negativity as a measure of average non-Gaussianity in Sec.~\ref{sec:xWLN}, maximize it in the case of a single PNRD in Sec.~\ref{sec:MaximumxWLN}, and discuss classes of states suited for single-mode Gaussian feed-forwards in Sec.~\ref{sec:SingleModeFF}.
Finally, we extend these results to the case of $k>1$ and demonstrate how multi-mode feed-forward compatibility can be introduced to general setups in Sec.~\ref{sec:GeneralMultiModeFF}.

\subsection{Expected Wigner logarithmic negativity}\label{sec:xWLN}

Given the Wigner function of a quantum state $W_\rho$, the Wigner logarithmic negativity (WLN) is defined as
\begin{align}
    \text{WLN}(\rho)=\log\left(\int d\vec{q}\left|W_\rho(\vec{q})\right|\right).
\end{align}
The WLN is an additive monotone of non-Gaussianity, i.e. it vanishes for Gaussian states and cannot increase under Gaussian operations or channels \cite{albarelliResourceTheoryQuantum2018,takagiConvexResourceTheory2018}. Notably, this also holds on average for Gaussian measurements, but can be broken by post-selection. As the WLN is additive, it allows for an easy comparison of different non-Gaussian resources by providing a lower bound on conversion rates.
Since the GBS-like setup is typically used as the only source of non-Gaussianity, a high initial WLN is an essential characteristic of any given output state, especially when realistic experimental noise is diminishing non-Gaussian features.

Apart from evaluating specific post-selected output states, the WLN can also be used to assess a full setup without post-selection by introducing the expected Wigner logarithmic negativity (xWLN):
\begin{align}
    \text{xWLN}\left(\Sigma, \vec{\mu}\right)=\sum_{\vec{n}}P_{\vec{n}}\cdot\log\left(\int d\vec{q}\left|W_{\Sigma, \vec{\mu}, \vec{n}}(\vec{q})\right|\right)
\end{align}
It quantifies the rate of non-Gaussianity a setup can produce and thereby provides a straightforward lower bound on the amount of (spatial or temporal) copies required to deterministically generate a given target state.
As a point of orientation, a deterministic source of GKP qunaught states \cite{walsheContinuousvariableGateTeleportation2020} above the squeezing threshold of 9.75 dB was shown to facilitate fault-tolerant and universal quantum computing \cite{aghaeeradScalingNetworkingModular2025,ostergaardOctoRailLatticeFourdimensional2025,baragiolaAllGaussianUniversalityFault2019a} and would require a WLN of 0.68 per state.

The choice of xWLN as figure of merit is convenient, given that the presented framework of Gaussian blurring, scaling, and filtering has a straightforward effect on the negative regions of the Wigner function and can thus provide some intuitive understanding of the resourcefulness of a given PSG state.
Furthermore, the theoretical description allows for an efficient numerical implementation of the xWLN: As the phase-space area of a Fock state scales linearly with $n$, the Gaussian convolution can be performed using a fast Fourier transform, and scaling and displacement can be done analytically, the contribution of large Fock states $n>150$ can be included without difficulty. This becomes necessary when high offline squeezing or displacements are considered.

\subsection{Maximizing the xWLN with a single PNRD}\label{sec:MaximumxWLN}

\begin{figure}[t]
    \centering
    \includegraphics[width=\linewidth]{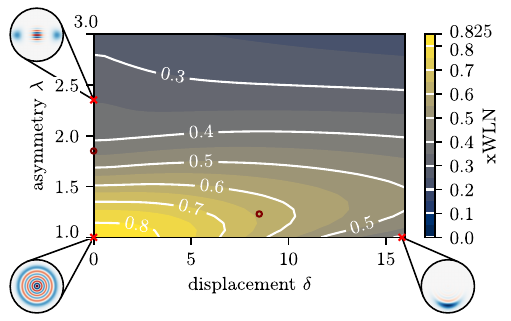}
    \caption{Expected Wigner logarithmic negativity (xWLN) for a GBS-like setup with a single PNRD, 15 dB offline squeezing, and no loss. The standard parameters for generating Fock-, cat-, and cubic phase-like states are marked with red crosses, while the custom choices shown in Figs.~\ref{fig:CatLike} and \ref{fig:CpsLike} are indicated by red circles.}
    \label{fig:Scan_ideal}
\end{figure}

The most common GBS-like setup involves only one photon detector $k=1$ and is known to produce cat-like as well as cubic phase-like states \cite{gottesmanEncodingQubitOscillator2001,takaseGenerationOpticalSchrodinger2021}. Both setups are similar, in that they allow Gaussian feed-forward operations for most measurement outcomes offering high success rates while being heavily affected by loss.
Beyond that, only Fock-like states as well as mixtures of these three output variants can be generated \cite{hanamuraStellarRankControl2025}.
In fact, we find that, besides the amount of offline squeezing, two real parameters suffice to describe any relevant PSG state for $k=1$: 
\\
First, we can fix $m=1$, as choosing $m>1$ only adds Gaussian elements which cannot increase the xWLN. Without loss, any symplectic $S_B$ as well as a final displacement also leave the xWLN unchanged and we can set $\vec{\beta}=0$ as well as $S_B=S_A^{-1}$ which minimizes the needed offline squeezing.
The matrix $S_A$ can be chosen diagonal due to the rotational symmetry of Fock states and the fact that a remaining rotation can be absorbed into $S_B$ using that for any symplectic matrix $S\in\text{Sp}(2,\mathbb{R})$ it is
\begin{align}
    Z_2SZ_2\in\text{Sp}(2,\mathbb{R}).
\end{align}
Finally, the displacement $\vec{\gamma}$ leaves two degrees of freedom. However, a displacement that maximizes the xWLN has to be perpendicular to the stronger blurring axis in order to highlight regions of Wigner negativity. Therefore, we are left with an amount of offline squeezing (alternatively, the two-mode squeezing $\xi_1$ could be fixed) as well as a choice of
\begin{align}
    S_A=
    \begin{bmatrix}
        \lambda & \\
         & \tfrac{1}{\lambda}
    \end{bmatrix}&&\text{and}&&
    \vec{\gamma}=
    \begin{bmatrix}
        0 \\
        \delta
    \end{bmatrix},
\end{align}
where $\lambda\geq1$ and $\delta\geq0$. While $\lambda$ describes the asymmetry needed to generate cat-like states, $\delta$ gives the displacement required to create cubic phase-like states. The resulting two-dimensional parameter space is shown in Fig.~\ref{fig:Scan_ideal}.

\begin{figure}[t]
    \centering
    \includegraphics[width=\linewidth]{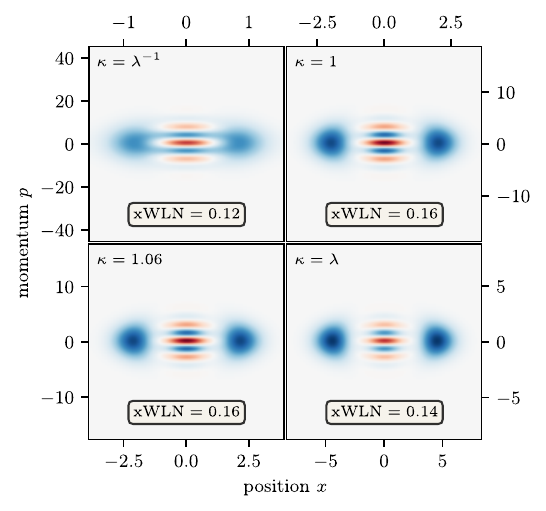}
    \caption{Impact of the choice of $\kappa$ on loss performance. Shown are a generated cat-like state as well as the total xWLN of the setup of Eq.~\eqref{eq:CatLike} with 15 dB offline squeezing, a loss of $\eta=0.95$, and the measurement result $n_1=5$. The four values of $\kappa$ represent the optimal choice for no loss ($\kappa=\lambda^{-1}$), the identity ($\kappa=1$), the optimal case ($\kappa=1.06$), and the unsqueezed cat ($\kappa=\lambda$).}
    \label{fig:OptimalB_cat}
\end{figure}

In case of loss, the symplectic matrix $S_B$ does have an effect on the xWLN. After aligning $S_B$ with $S_A$ and $\vec{\gamma}$ this adds one parameter
\begin{align}
    S_B=
    \begin{bmatrix}
        \kappa & \\
         & \tfrac{1}{\kappa}
    \end{bmatrix},
\end{align}
where $\kappa>0$.
Yet, any $\kappa$ can be undone by a Gaussian operation later on and choosing $\kappa$ to maximize the xWLN of a given setup does not affect the type of output states. Therefore, calculations of the xWLN will implicitly be using the optimal $\kappa$. Figure~\ref{fig:OptimalB_cat} shows the impact of different $\kappa$ in case of a cat-like state with an offline squeezing of 15 dB and a loss of $\eta=0.95$.
We find that the common choices of $\kappa=\lambda^{-1}$, which is optimal without loss, and $\kappa=\lambda$, which approximates a true cat state, do indeed become suboptimal once loss is present. Optimal $\kappa$ values can be found in App.~\ref{sec:AppendixFigures}.
The effect of loss can also be mitigated by relying more heavily on low photon numbers. Remarkably, it was demonstrated in \cite{hanamuraStellarRankControl2025} that the photon number distribution of a given setup can be adjusted without affecting its output states. A related setup with reduced offline squeezing is thereby found by introducing a damping operation before the detector. In case of post-selection, this can be used to boost the setup's success rate. Once feed-forward operations are included, however, this increases the likelihood of detecting vacuum and is mostly detrimental to the overall xWLN. The amount of squeezing will therefore not be optimised for different levels of loss but rather remain as specified.

\begin{figure}[t]
    \centering
    \includegraphics[width=\linewidth]{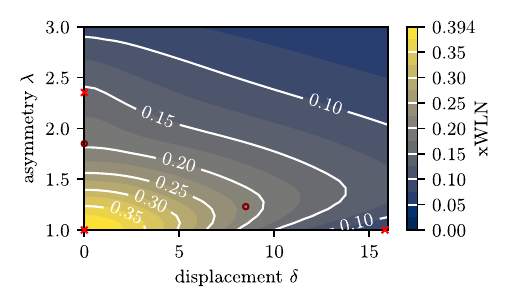}
    \caption{Expected Wigner logarithmic negativity (xWLN) for a GBS-like setup with a single PNRD, 15 dB offline squeezing, and a photon loss of $\eta=0.95$. The standard parameters for generating Fock-, cat-, and cubic phase-like states are marked with red crosses, while the custom choices shown in Figs.~\ref{fig:CatLike} and \ref{fig:CpsLike} are indicated by red circles.}
    \label{fig:Scan_95}
\end{figure}

Provided with the available amount of offline squeezing as well as the experienced photon loss of an experimental setup, we can now map the xWLN for the full two-dimensional parameter space of relevant PSG states spanned by $\lambda$ and $\delta$.
For an offline squeezing of 15 dB, the xWLN for lossless setups is presented in Fig.~\ref{fig:Scan_ideal}, while Fig.~\ref{fig:Scan_95} shows the same setups after a loss of $\eta=0.95$. Corresponding plots for sole detector and sole output loss can be found in App.~\ref{sec:AppendixFigures}.
We find that setups generating Fock-like states generally exhibit the highest xWLN even under lossy conditions. Increasing either the asymmetry or the displacement significantly diminishes the measured xWLN. Along the axes, the generation of cat-like states produces more xWLN than the generation of cubic phase-like states when loss is present, but slightly less without. In general, the Fock- and cubic phase-like state generation is less affected by detector losses, while cat-like state generation can tolerate more output losses (see Fig.~\ref{fig:Scan_B}).
Furthermore, both figures show that sticking to the axes is not necessarily optimal. Most notably, adding some asymmetry to setups with high displacement significantly increases their xWLN as well as loss tolerance.

\subsection{Single-mode feed-forward operations}\label{sec:SingleModeFF}
While the xWLN gives a good indication of a setup's potential, it cannot differentiate between usable and unusable Wigner negativity. Additionally required are feed-forward operations that can bring its Wigner negativity into a common form and allow it to be used independent of measurement outcomes. Therefore, we take a closer look at the different classes of states
as well as two distinct feed-forward strategies.

\begin{figure}[t]
    \centering
    \includegraphics[width=\linewidth]{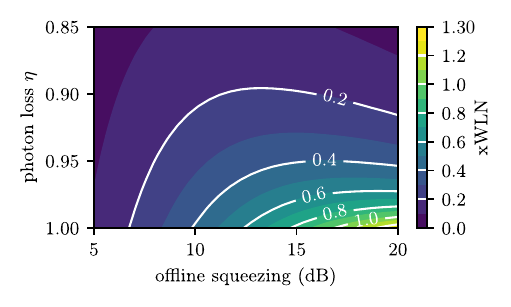}
    \caption{Expected Wigner logarithmic negativity (xWLN) of the GBS-like setup of Eq.~\eqref{eq:FockLike} generating Fock-like states for different amounts of offline squeezing and photon loss. The used $\kappa$ values can be found in App.~\ref{sec:AppendixFigures}.}
    \label{fig:xWLN_Fock}
\end{figure}

\subsubsection{Fock-like states}\label{sec:FockLikeStates}
The natural parameters to generate Fock-like states are
\begin{align}\label{eq:FockLike}
    \lambda=1&&\text{and}&&\delta=0
\end{align}
retaining the rotational symmetry of the initial Fock states. However, any states that exhibit the typical rings of Wigner negativity belong to this class. Precisely these rings prohibit any type of feed-forward, as their number and radii depend on the measured photon number and cannot be meaningfully aligned using Gaussian operations. On the other hand, they exhibit the highest xWLN of all setups. Therefore, they establish a straight-forward lower bound of the amount of resources needed to generate a desired state under realistic conditions. Figure~\ref{fig:xWLN_Fock} shows the xWLN obtained for different amounts of offline squeezing and photon loss. The absolute minimum number of identical setups needed to generate a specific state at a quasi-deterministic rate is then given by the ratio of the state's WLN and the setup's xWLN. Interestingly, the results indicate that there exists an optimal amount of offline squeezing for a given photon loss.

Presumably, the resource bound also extends to larger setups with $k>1$: As the xWLN is maximal for $m=k>1$ and its calculation quickly becomes computationally intractable due to the $2m$-dimensional Wigner integral in combination with an exponentially growing number of photon patterns, we cannot verify this directly.
Nevertheless, setting all two-mode squeezing $\xi_1=\xi_2=...=\xi_\text{max}$ to the maximal available amount, we can again reduce the number of parameters using the same arguments as in the case of $k=m=1$. This leaves us with $k$ independent two-mode setups just connected by an orthogonal symplectic matrix acting on the initial $W_{\vec{n}}$. As its effect vanishes for vanishing asymmetry, which in turn was shown to introduce additional blurring and reduce negativity, this strongly suggests that the maximal xWLN of a setup with $k>1$ is indeed $k$ times the maximal xWLN of a setup with $k=1$.

\subsubsection{Cat-like states}\label{sec:CatLikeStates}

\begin{figure}[t]
    \centering
    \includegraphics[width=\linewidth]{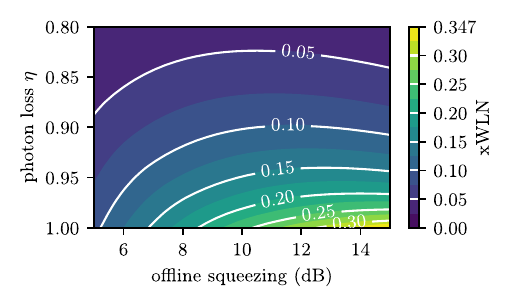}
    \caption{Expected Wigner logarithmic negativity (xWLN) of the GBS-like setup of Eq.~\eqref{eq:CatLike} generating cat-like states for different amounts of offline squeezing and photon loss. The used $\kappa$ values can be found in App.~\ref{sec:AppendixFigures}.}
    \label{fig:xWLN_Cat}
\end{figure}

Following \cite{takaseGenerationOpticalSchrodinger2021}, the most efficient setup to generate cat-like states is given by
\begin{align}
    \lambda=\sqrt{\cosh{\xi_1}}&&\text{and}&&\delta=0
    \label{eq:CatLike}
\end{align}
resulting in $A_{11}=1$. Figure~\ref{fig:xWLN_Cat} shows the xWLN of this setup for varying squeezing and loss. Unlike for setups generating Fock- and cubic phase-like states, these results are significantly affected by the optimal choice of $\kappa$, resulting in output states that are well protected against the offline/output loss $\eta_B$. Corresponding optimal $\kappa$ values can be found in App.~\ref{sec:AppendixFigures}.

\begin{figure}[t]
    \centering
    \includegraphics[width=\linewidth]{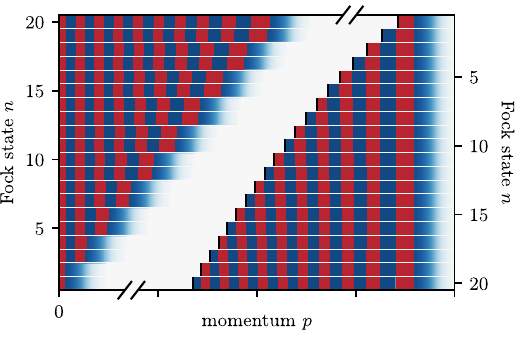}
    \caption{Effect of the two distinctive single-mode feed-forwards on the underlying Fock Wigner functions. The latter are represented by their positive (blue) and negative (red) sections along the $p$-axis. The cat-like feed-forward from Eq.~\eqref{eq:FFcat} aligns the centre stripes (left), while the cubic phase-like feed-forward from Eq.~\eqref{eq:FFcps} aligns the outermost arcs (right). For reasons of symmetry (left) and their vanishing contribution (right), the sections to the left of the origin are truncated.}
    \label{fig:FeedForward}
\end{figure}
\begin{figure}[t]
    \centering
    \includegraphics[width=\linewidth]{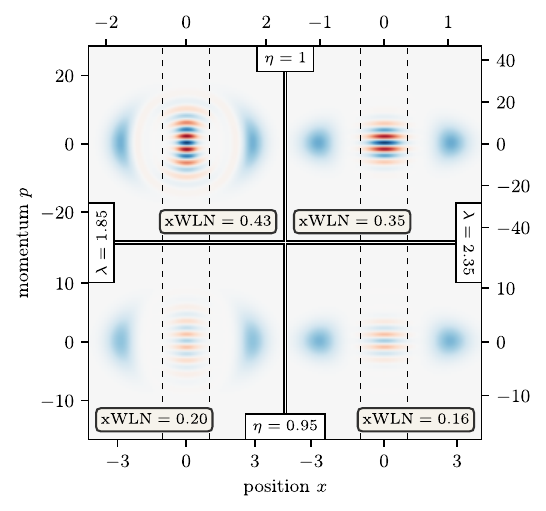}
    \caption{Comparison of the discussed custom (left) and standard (right) setups for generating cat-like states. Shown are the output and total xWLN of both setups for 15 dB offline squeezing, the measurement result $n_1=8$, and losses of $\eta=1$ as well as $\eta=0.95$. The marked region represents the two-sigma interval of the Gaussian blur and can be used to indicate a setup's feed-forward compatibility.}
    \label{fig:CatLike}
\end{figure}

Cat-like states distinguish themselves by the positive and negative stripes found within the centre of the Wigner function caused by the asymmetric blurring. Furthermore, the asymmetric filtering suppresses the positive outer ring in the direction perpendicular to the stripes. In combination, these two effects enable suited feed-forward operations and facilitate near-deterministic generation rates. Specifically, it is known that the stripe spacing scales roughly with $1/\sqrt{n}$, while the distance between the two outer ring remnants scales with $\sqrt{n}$. The appropriate squeezing operation together with a small displacement aligning positive and negative stripes therefore brings all output states into the same form. This also includes states with no or few stripes resulting from measuring no or few photons. Note that the exact Gaussian operation should be chosen to enact the necessary scaling and displacement on the Fock state directly, so that
\begin{align}
    C\mapsto \left[
    \begin{smallmatrix}
        \sqrt{n} &  \\
         & 1/\sqrt{n}
    \end{smallmatrix}
    \right]\cdot C, && \vec{\gamma}\mapsto \vec{\gamma} + \tfrac{1+(-1)^{n}}{2}\left[
    \begin{smallmatrix}
        0 \\
        1.53/\sqrt{n}
    \end{smallmatrix}
    \right].\label{eq:FFcat}
\end{align}
The effect of this feed-forward acting on the set of underlying Fock Wigner functions is depicted on the left-hand side of Fig.~\ref{fig:FeedForward}.

In terms of xWLN, the setup of Eq.~\eqref{eq:CatLike} can be improved by reducing the asymmetry $\lambda$, whereas the introduction of a displacement is found to have no major upside. The optimal setup for an application relying on the generation of cat states will therefore be a compromise between its xWLN and feed-forward compatibility.
In order to get an indication about the latter, we calculate the percentage of a setup's xWLN that lies within the narrow band depicted in Fig.~\ref{fig:CatLike}. The width of the band was chosen to be the two-sigma interval of the Gaussian blur $4A_{11}=4$ acting on the initial Fock state. For an offline squeezing of 15 dB, we find that a setup with $\lambda=1.85$ thereby still retains 95\% of its non-Gaussianity compared to 98\% for the setup which fulfils Eq.~\eqref{eq:CatLike}, yet exhibits a 24\% higher total xWLN. This becomes even more pronounced when loss is involved, as $A_{11}\stackrel{\eta\rightarrow0}{\longrightarrow}1$. For a loss of $\eta=0.95$, both setups show a feed-forward compatibility of 99\% while the more symmetric setup still possesses a 21\% higher xWLN. Cat-like states generated by the two setups with and without loss and a measurement result of $n_1=8$ are shown in Fig.~\ref{fig:CatLike}.

Besides highlighting the consequences of deviating from Eq.~\eqref{eq:CatLike}, these exemplary Wigner functions also provide some information about the effect of loss:
While a state's Wigner negativity is generally diminished by additional blurring, the outlines of its negative regions also change. Remarkably, this can be easily accounted for when designing a setup.
As the matrix $A_\text{loss}$ remains positive definite and symmetric, its Williamson decomposition \cite{williamsonAlgebraicProblemConcerning1936a} returns the related pure symplectic matrix $S_A'$ and reduced two-mode squeezing $C_h'$. Since both filtering and scaling conserve the sign of the Wigner function, this pure setup is equivalent to the lossy one up to a different filter $B'$ and the potential scaling of $\sqrt{\eta_B/\eta_A}$. Consequently, it is possible to optimise a pure setup with appropriately low squeezing and subsequently adjust it to the level of loss, rather than to optimise the lossy setup directly.
In case of the more (less) symmetric cat-like state of Fig.~\ref{fig:CatLike}, the related pure setup is found to have an offline squeezing of 10.6 dB (10.4 dB) and offers an xWLN of 0.28 (0.24). For a direct comparison of their generated Wigner functions see App.~\ref{sec:AppendixFigures}.
Notably, this relation becomes approximate in the case of $k>m$, as the squeezing $C_h'$ is then partially fixed.
Moreover, any displacements need to be adapted accordingly.

\subsubsection{Cubic phase-like states}\label{sec:CpsLikeStates}

\begin{figure}[t]
    \centering
    \includegraphics[width=\linewidth]{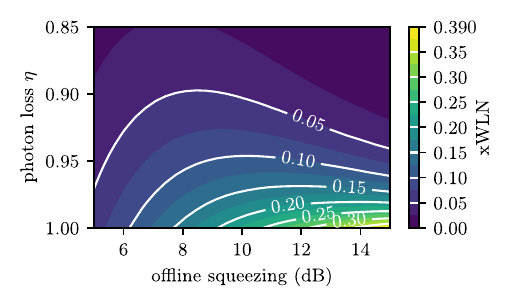}
    \caption{Expected Wigner logarithmic negativity (xWLN) of the GBS-like setup of Eq.~\eqref{eq:CpsLike} generating cubic phase-like states for different amounts of offline squeezing and photon loss. The used $\kappa$ values can be found in App.~\ref{sec:AppendixFigures}.}
    \label{fig:xWLN_Cps}
\end{figure}

The near-deterministic generation of cubic phase-like states is described in \cite{gottesmanEncodingQubitOscillator2001} requiring a two-mode squeezed state and sufficiently high displacement.
Hereto, we use
\begin{align}
    \lambda=1&&\text{and}&&\delta=\cosh{\xi_1},
    \label{eq:CpsLike}
\end{align}
i.e.\ the one-sigma interval of the Gaussian filter $B_{11}$.
Figure~\ref{fig:xWLN_Cps} shows the xWLN of this setup for varying squeezing and loss. With an optimal $\kappa=\lambda=1$, the interplay of photon loss and squeezing is particularly straightforward. From Eq.~\eqref{eq:LossApproximation} we obtain a direct relation of the blur caused by low squeezing and the different loss sources,
\begin{align}
    A_\text{loss} \approx{}&\eta_A \left(\left(\cosh\xi_1\right)^{-1} + \left(\eta_A^{-1}-1\right) + \left(\eta_B^{-1}-1\right)\right)I_2.
\end{align}
Consequently, setups following Eq.~\eqref{eq:CpsLike} protect their regions of Wigner negativity and as a result their overall xWLN better against detector than output losses.

In general, cubic phase-like states are characterised by a Wigner function consisting of alternating circular arcs, achieved by focusing the Gaussian filter on a section of the positive outer ring. While long arcs with different radii are difficult to align using Gaussian operations, shorter arcs retain less negativity. In this regard, we find preserving a quarter of the outer ring to be a good compromise. Then, the one central as well as two end points of an arc of radius $r_1$ can be made to lie on a larger arc of radius $r_2>r_1$ by applying a squeezing of $s(r_2/r_1)$, where $s(r)$ is the only real solution to the equation
\begin{align}
    s(r)^4 - 4\left(1-\tfrac{1}{\sqrt{2}}\right)r\cdot s(r) + 2\left(1-\tfrac{1}{\sqrt{2}}\right)^2 = 0
\end{align}
that is greater than one.
With the radius of the Wigner function of a Fock state approximately proportional to $\sqrt{n}$, we find that this necessary squeezing can be well approximated as being proportional to the fifth root of $n$,
\begin{align}
    s(\sqrt{n})\appropto \sqrt[5]{n}.
\end{align}
On the other hand, the width of the outer rings of Fock states happens to roughly scale with $1/\sqrt[5]{n}$. Together with an appropriate displacement aligning the outermost negative arcs, this thus allows the cubic phase-like outputs of a given setup to be brought into one common form, enabling near-deterministic generation rates.
The effect of this feed-forward is depicted on the right-hand side of Fig.~\ref{fig:FeedForward} when enacted on the set of underlying Fock Wigner functions. Specifically, it maps
\begin{align}
    C\mapsto \left[
    \begin{smallmatrix}
        \sqrt[5]{n} &  \\
         & 1/\sqrt[5]{n}
    \end{smallmatrix}
    \right]\cdot C &&\text{and}&& \vec{\gamma}\mapsto \vec{\gamma} - \left[
    \begin{smallmatrix}
        0 \\
        \sqrt{z_n}
    \end{smallmatrix}
    \right],\label{eq:FFcps}
\end{align}
where $z_n$ is the outermost zero of the $n$-th Laguerre polynomial.

\begin{figure}[t]
    \centering
    \includegraphics[width=\linewidth]{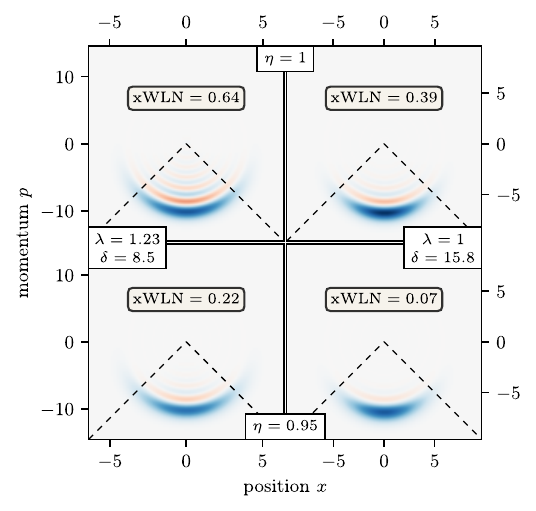}
    \caption{Comparison of the discussed custom (left) and standard (right) setups for generating cubic phase-like states. Shown are the output and total xWLN of both setups for 15 dB offline squeezing, the measurement result $n_1=12$, and losses of $\eta=1$ as well as $\eta=0.95$. The marked region represents the bottom quarter of the initial Fock state and can be used to indicate a setup's feed-forward compatibility.}
    \label{fig:CpsLike}
\end{figure}

In terms of xWLN, the setup of Eq.~\eqref{eq:CpsLike} can be improved by reducing the displacement $\delta$ as well as adding some asymmetry $\lambda$. In order to assess whether such setups are compatible with the described feed-forward scheme, we calculate the percentage of a setup's xWLN that lies within the bottom quarter of the initial Fock states as indicated in Fig.~\ref{fig:CpsLike}.
For an offline squeezing of 15 dB, a setup with $\delta=8.5$ and $\lambda=1.23$ thereby retains 93\% of its non-Gaussian features compared to 99\% for the setup following Eq.~\eqref{eq:CpsLike}, but exhibits a 64\% higher total xWLN. With a loss of $\eta=0.95$, its feed-forward compatibility increases to 95\% while a significantly better loss protection leads to a 226\% higher total xWLN compared to the setup of Eq.~\eqref{eq:CpsLike}. These results highlight that setups involving displacements can be significantly improved by finding the right amount of asymmetry $\lambda$ along with the optimal $\kappa$. Cubic phase-like states generated by the two setups with and without loss and a measurement result of $n_1=12$ are shown in Fig.~\ref{fig:CpsLike}.

We expect the two presented feed-forward variants to be peerless, as inclusion of the centre of the Fock Wigner function requires the scaling of $\sqrt{n}$ as well as stripes, while the inclusion of the outer ring necessitates a scaling of $\sqrt[5]{n}$ as well as circular arcs. In between, any focus point would require sufficiently high $n$ along with a narrow filter envelope making it unlikely to succeed.
Moreover, any near-deterministic multi-mode feed-forward needs the ability to correct each mode individually and will consequently be based on a combination of the two presented variants.

\subsection{General multi-mode feed-forward}\label{sec:GeneralMultiModeFF}
Whenever more complex non-Gaussian states are needed, a larger setup with $k>1$ becomes necessary.
Here, there are two valid pathways: One can choose a \textit{compact} setup with $m<k$ relying on post-selection \cite{suConversionGaussianStates2019a,larsenIntegratedPhotonicSource2025b,takaseGottesmanKitaevPreskillQubitSynthesizer2023} or a \textit{composite} setup merging multiple cat- and cubic phase-like states by Gaussian measurements and operations \cite{takaseGenerationFlyingLogical2024,konnoPropagatingGottesmanKitaevPreskillStates2024,budingerAllopticalQuantumComputing2024a}.
\textit{Composite} setups, also known as breeding protocols, provide full feed-forward compatibility at the cost of a limited reachable state space. \textit{Compact} setups, on the other hand, are general, but come with a vast parameter space and no meaningful feed-forward operations. As a result, they are typically optimised for a small set of photon patterns and exhibit vanishing success rates. Once a \textit{composite} setup is given, an equivalent \textit{compact} version can be obtained using the Gaussian formalism \cite{weedbrookGaussianQuantumInformation2012,braskGaussianStatesOperations2022}, however, the reverse is generally not true.
In the following, we introduce a \textit{hybrid} variant, which produces the same output states as any given \textit{compact} setup, while providing feed-forward compatibility. Furthermore, we discuss how to find a similar \textit{composite} variant and compare their performance in the case of GKP-like state generation.

\subsubsection{Setup variants}\label{sec:SetupVariants}

\begin{figure}[t]
    \centering
    \includegraphics[width=\linewidth]{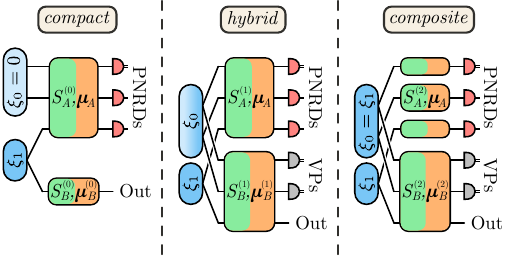}
    \caption{Schematic depiction of the three different setup variants generating PSG states. In all cases, an array of two-mode squeezed states with a squeezing of $\xi_1$ and $0\leq\xi_0\leq\xi_1$ is subjected to the Gaussian transformations $\left(S_A, \vec{\mu}_A\right)$ and $\left(S_B, \vec{\mu}_B\right)$ and subsequently measured by photon-number resolving detectors (PNRDs) as well as vacuum projections (VPs). The latter can be replaced by Gaussian measurements requiring an additional step of post-selection and feed-forward. The setups are characterized by a non-entangling $S_B^{(0)}$ and $\xi_0=0$ (\textit{compact}) as well as a non-entangling $S_A^{(2)}$ (\textit{composite}). The \textit{hybrid} setup is designed to retain the \textit{compact} $S_A^{(1)}=S_A^{(0)}$ besides introducing feed-forward compatibility through an adaptable $S_B^{(1)}$. Note that the depicted order of modes differs from the mathematical description.}
    \label{fig:ThreeSetups}
\end{figure}

Let us consider a pure \textit{compact} setup with $m<k$ characterised by
\begin{align}
    \left(\vec{\xi}^{(0)}, S_A^{(0)}, S_B^{(0)}, \vec{\beta}^{(0)}, \vec{\gamma}^{(0)}\right).
    \label{eq:FFvariant1}
\end{align}
With all relevant entanglement comprised in $S_A^{(0)}$, it offers the smallest possible experimental footprint, but relies on post-selection as $S_B^{(0)}$ can only act directly on the output state.
We can add $(k-m)$ vacuum modes which are measured by $(k-m)$ on-off detectors (i.e.\ vacuum projections) to obtain an equivalent setup with $m'=k$. Moreover, introducing some two-mode squeezing $\xi_{m+1}^{(1)}, ..., \xi_{k}^{(1)}$ between pairs of vacuum modes does not change its output states as long as vacuum is post-selected. With the assumption that $\xi_1^{(0)}=\xi_2^{(0)}=...=\xi_m^{(0)}$ as well as $\xi_{m+1}^{(1)}=...=\xi_k^{(1)}\equiv\xi_0$, this \textit{hybrid} setup is given by
\begin{multline}
    \left(\vec{\xi}^{(1)}=\left[
    \begin{smallmatrix}
        \vec{\xi}^{(0)} \\
        \xi_0 \vec{1}
    \end{smallmatrix}
    \right], S_A^{(1)}=S_A^{(0)}, S_B^{(1)}=\left[
    \begin{smallmatrix}
        S_B^{(0)} &  \\
         & I
    \end{smallmatrix}
    \right],\right.\\
    \left.\vec{\beta}^{(1)}=\left[
    \begin{smallmatrix}
        \vec{\beta}^{(0)} \\
        \vec{0}
    \end{smallmatrix}
    \right], \gamma^{(1)}=\gamma^{(0)}\right)
    \label{eq:FFvariant2}
\end{multline}
as illustrated in Fig.~\ref{fig:ThreeSetups}.
Compared to the \textit{compact} variant obtained for $\xi_0=0$, the \textit{hybrid} setup introduces an additional post-selection with a success probability of $1/\cosh\left(\xi_0/2\right)$ per vacuum projection, but enables the implementation of feed-forward operations:
For $\xi_0=\xi_1^{(0)}$, a multi-mode operation exists that maps
\begin{align}\label{eq:MultiModeFF}
    C^{(1)}\mapsto \bigoplus_{j=1}^k S_{\text{FF}, j}\cdot C^{(1)}, && \vec{\gamma}^{(1)}\mapsto \vec{\gamma}^{(1)} + \bigoplus_{j=1}^k \vec{d}_{\text{FF}, j},
\end{align}
where $S_{\text{FF}, j}$ and $\vec{d}_{\text{FF}, j}$ denote single-mode feed-forwards. This still holds approximately for varying but sufficiently high two-mode squeezing, so that $\tanh \xi_1^{(1)}, ..., \tanh \xi_{2k}^{(1)}\approx1$, but can additionally lower the success probability of the vacuum projections. Consequently, the parameter  $0\leq\xi_0\leq\xi_1^{(0)}$ acts as a dial between the \textit{compact} and \textit{hybrid} setup variant and mediates between higher individual success rates and better feed-forward performance.

\begin{figure}[t]
    \centering
    \includegraphics[width=\linewidth]{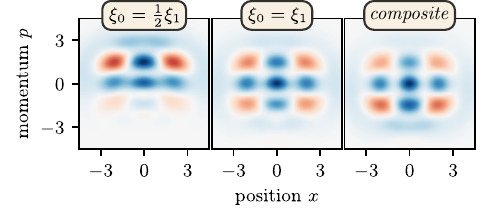}
    \caption{Feed-forward performance of \textit{hybrid} and \textit{composite} GKP-like state generation. Shown are the output states obtained for $n_1=2$, $n_2=5$, and a successful vacuum projection using \textit{hybrid} setups with $\xi_0=\frac{1}{2}\xi_1$ and $\xi_0=\xi_1$, as well as the \textit{composite} variant.}
    \label{fig:GKPffComparison}
\end{figure}

Even for $\xi_0=\xi_1^{(0)}$, this \textit{hybrid} feed-forward is not ideal. How effective it can be depends on how well it commutes with the unaffected blurring introduced by $A^{(1)}$. Here, a diagonalization reveals the blurring axes as well as their strengths. Whenever a relevant axis mixes two or more modes, the feed-forward will cause it to rotate in phase space and misalign. This can be avoided by choosing a diagonal $A^{(1)}$ through moving all mode-entangling from $S_A^{(1)}$ to $S_B^{(1)}$.
The resulting setup,
\begin{align}
    \left(\vec{\xi}^{(2)}=\vec{\xi}^{(1)}, S_A^{(2)}, S_B^{(2)}, \vec{\beta}^{(2)}=\vec{\beta}^{(1)}, \vec{\gamma}^{(2)}=\vec{\gamma}^{(1)}\right),
    \label{eq:FFvariant3}
\end{align}
where $S_A^{(2)}$ is diagonal, is exactly the \textit{composite} variant, consisting of $k$ two-mode devices and subsequent Gaussian operations and measurements.
With all relevant entanglement included in $S_B^{(2)}$, it offers full compatibility with any given feed-forward operation, along with further lowered success rates of the vacuum projections. A comparison of the feed-forward quality of the \textit{hybrid} and \textit{composite} setup generating GKP-like states is depicted in Fig.~\ref{fig:GKPffComparison}.

When a \textit{hybrid} setup is given, finding a closely related \textit{composite} variant comes down to finding an appropriate pair of $S_A^{(2)}$ and $S_B^{(2)}$ that approximately conserves the effective blurring $A^{(0)}$.
For sufficiently high squeezing, this total blurring can be estimated by disregarding the envelope $B^{(1/2)}$ leading to the relation
\begin{align}\begin{split}
    A^{(0)} &\approx A^{(1)} + C^{(1)}\left[
    \begin{smallmatrix}
        0 &  \\
         & I_{2k-2m}
    \end{smallmatrix}
    \right]C^{(1)T}\\
    &\approx A^{(2)} + C^{(2)}\left[
    \begin{smallmatrix}
        0 &  \\
         & I_{2k-2m}
    \end{smallmatrix}
    \right]C^{(2)T},
    \label{eq:BlurTotal}
\end{split}\end{align}
which can serve as a starting point for the search.
In the common case of $m=1$ along with high squeezing $\xi_0=\xi_1^{(0)}$, the matrix $A^{(1)}$ is diagonalized by an orthogonal symplectic matrix and can thus have up to $k$ non-vanishing blurring axes. A sufficient condition is then that their strengths can be levelled by moving blurring to and from the $(k-1)$ vacuum projections, leaving them disentangled and allowing to gradually move the entanglement from $S_A^{(1)}$ to $S_B^{(1)}$.

Finally, the effect of the added post-selection can be mitigated by replacing the vacuum projections with Gaussian measurements, namely heterodyne detections. This way, all measurements succeed, but shift the displacement $\vec{\beta}^{(1/2)}$ and thus $\vec{\gamma}^{(1/2)} = \vec{\mu}_A - C^{(1/2)}\vec{\beta}^{(1/2)}$ depending on their outcome. Consequently, the post-selection of a range of measurement results will still be necessary. How leniently this is administered decides both the new success rates as well as the deviation of the generated states from those obtained by vacuum projections. An advantage of using heterodyne measurements is the option to include active Gaussian operations from $S_B^{(1/2)}$ as well as the feed-forward into a general-dyne measurement which can be implemented by passive optical components.

\begin{figure}[t]
    \centering
    \includegraphics[width=\linewidth]{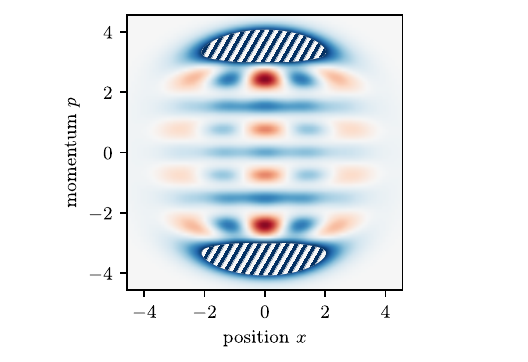}
    \caption{Underlying Gaussian blur of a GKP-like output state. The filtering and scaling of the state obtained for an offline squeezing of 20 dB, $k=3$, and $n_1=n_2=n_3=4$ are omitted to highlight the two perpendicular blurring axes. The resulting peak layout consists of $n_1$ rows and $k$ columns of negative peaks. Dashed regions lie outside of the chosen colour range.}
    \label{fig:GKPexample}
\end{figure}

\begin{figure*}[t]
    \centering
    \includegraphics[width=\linewidth]{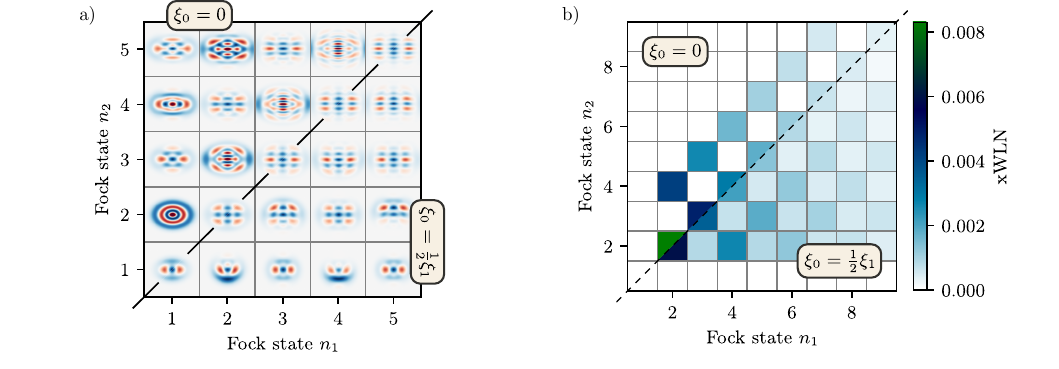}
    \caption{Comparison of the \textit{compact} ($\xi_0=0$) and \textit{hybrid} ($\xi_0=\frac{1}{2}\xi_1$) GKP-like state generation. Shown are a) the corresponding output state and b) the post-selected xWLN for different photon patterns $(n_1, n_2)$. The permutation $(n_2, n_1)$ leaves the results unaffected and is omitted. For $n_1=n_2$, both setups produce the same state but a higher $\xi_0$ leads to a lower success rate.}
    \label{fig:GKPcompactVsHybrid}
\end{figure*}

\subsubsection{GKP-like states}\label{sec:GKPLikeStates}

Common target states requiring $k>1$ are GKP-like states \cite{gottesmanEncodingQubitOscillator2001}, characterised by a rectangular grid of positive and negative Gaussian peaks in phase space. In their repeating two-by-two unit cell exactly one peak is negative, the choice of which decides the logical Pauli eigenstate. Generation schemes based on breeding cat-like states \cite{takaseGenerationFlyingLogical2024,solodovnikovaLossToleranceCat2025,aghaeeradScalingNetworkingModular2025} as well as post-selected photon subtraction \cite{larsenIntegratedPhotonicSource2025b,takaseGottesmanKitaevPreskillQubitSynthesizer2023,aghaeeradScalingNetworkingModular2025} are well researched. Notably, the \textit{compact} photon subtraction setups are derived directly from the \textit{composite} breeding protocols, allowing for a straightforward comparison.
\paragraph*{Compact setup.}\label{sec:CompactSetup}
For the generation of GKP qunaught states, which showcase off-axis negative peaks on a square grid (see Fig.~\ref{fig:GKPffComparison}), the \textit{compact} setup for $k$ detectors is given by
\begin{align}
    A&= O_A D_A O_A^T, & B &= \left[
    \begin{smallmatrix}
        \cosh^2\xi_1 &  \\
         & 1
    \end{smallmatrix}
    \right], & C = \tfrac{\tanh\xi_1}{\sqrt{k}}\left[
    \begin{smallmatrix}
        Z_2\\
        Z_2\\
        \vdots
    \end{smallmatrix}
    \right],\nonumber\\
    \vec{\beta}&=\vec{0}_2,& \vec{\gamma}&=\vec{0}_{2k},
    \label{eq:GKPsetup}
\end{align}
where $D_A=\text{diag}\left(1, \frac{1}{\cosh^2\xi_1}; \cosh^2\xi_1, \frac{1}{\cosh^2\xi_1}, ...\right)$ and
\begin{align}
    O_A=\tfrac{1}{\sqrt{k}}\left[
    \begin{NiceMatrix}
        1&0&\cdot&0&\\
        0&1&0&\cdot&\\
        1&0&\cdot&0&\\
        0&1&0&\cdot&\\
        \multicolumn{2}{c}{$\vdots$}&&&\ddots
    \end{NiceMatrix}
    \right].
\end{align}
The exact entries of column three and onwards are insignificant as long as they do not mix $x$ and $p$ and $O_A$ remains orthonormal.
Relevant blurring happens both along the $x$-diagonal with strength 1 as well as in the perpendicular $x$-directions with strength $\cosh^2\xi_1$, leading to horizontal and vertical blue stripes across the Wigner function and the desired grid pattern. When measuring the same photon number $n$ at all $k$ detectors, there are simply $n$ rows and $k$ columns of negative peaks as highlighted in Fig.~\ref{fig:GKPexample} for an offline squeezing of 20 dB, $k=3$, $n=4$, and an omitted filtering and scaling of the output state.
Just like the cat-like states from Eq.~\eqref{eq:CatLike} it is based on, this setup is not optimal. The blurring $D_A$ can be reduced without noticeably altering the state, while in turn broadening the envelope $B$ of the filter and revealing additional peaks \cite{takaseGenerationFlyingLogical2024}. Moreover, the required amount of offline squeezing can be minimised by changing the squeezing of the output. For all following results, we will use an offline squeezing of 12.5 dB along with an asymmetry $\lambda=1.72$ and $\kappa=\lambda^{-1}$ of the underlying cat-like states leading to
\begin{align}
    D_A=\text{diag}{\textstyle\left(\frac{\lambda^2}{\cosh\xi_1}, \frac{\lambda^{-2}}{\cosh\xi_1}; \lambda^2\cosh\xi_1, \frac{\lambda^{-2}}{\cosh\xi_1}, ...\right)}
\end{align}
as well as
\begin{align}
    B = \cosh\xi_1\left[
    \begin{smallmatrix}
        \kappa^2 &  \\
         & 1/\kappa^2
    \end{smallmatrix}
    \right],&& C=\tfrac{\tanh\xi_1}{\sqrt{k}}\left[
    \begin{smallmatrix}
        Z_2\\
        Z_2\\
        \vdots
    \end{smallmatrix}
    \right]\cdot \left[
    \begin{smallmatrix}
        \lambda/\kappa &  \\
         & \kappa/\lambda
    \end{smallmatrix}
    \right].
\end{align}
For illustration purposes, however, all states are displayed after being scaled to match the case of $\kappa=\lambda$.
The output of this \textit{compact} setup for $k=2$ is shown at the top of Fig.~\ref{fig:GKPcompactVsHybrid}a. Noticeably, states obtained for even/odd $n_1=n_2$ share the same form. This is common, as identical feed-forward squeezing $S_{\text{FF}}$ on all modes commutes with any Gaussian operation and can be applied on the output state directly.
A setup optimized for $n_{1, \text{opt}}$ and $n_{2, \text{opt}}$ will thus perform well as long as $\frac{n_1}{n_2}=\frac{n_{1, \text{opt}}}{n_{2, \text{opt}}}$ making $n_{1, \text{opt}}=n_{2, \text{opt}}$ the natural choice. On the other hand, the feed-forward displacement $d_\text{FF}$ does not commute and causes the difference in outputs with even and odd $n_1=n_2$.

\begin{figure*}[t]
    \centering
    \includegraphics[width=\linewidth]{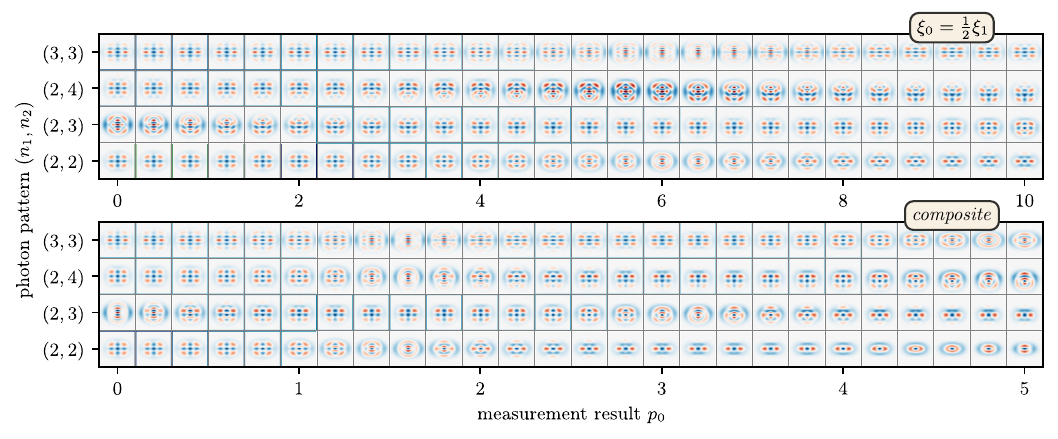}
    \caption{GKP-like output states of the \textit{hybrid} ($\xi_0=\frac{1}{2}\xi_1$) and \textit{composite} setups using homodyne detection for different measurement results $p_0$ and photon patterns $(n_1, n_2)$.}
    \label{fig:Homodyne_grid}
\end{figure*}

\paragraph*{Compact vs hybrid setup.}\label{sec:HybridSetup}
In order to evaluate the performance of the different setup variants, we consider the generation of a qunaught state with four or more negative peaks for $k=2$. 
Then, outputs with even $n_1=n_2$ just require the feed-forward squeezing of
\begin{align}
    S_\text{out}\left(n_1\right)=\left[\begin{smallmatrix}
        \sqrt{\pi/(kn_1)} &  \\
         & \sqrt{kn_1/\pi}
    \end{smallmatrix}\right]
\end{align}
to be aligned with the symmetric grid, while outputs with odd $n_1=n_2>1$ also need a displacement. Finally, the same holds for states with $(n_1 - n_2)=\pm2$. The xWLN distribution of the \textit{compact} setup resulting from this post-selection is shown at the top of Fig.~\ref{fig:GKPcompactVsHybrid}b for $n_1, n_2\leq 9$ and sums up to 0.041.
In comparison, the \textit{hybrid} setup with $\xi_0=\frac{1}{2}\xi_1$ and vacuum projection can limit the post-selection to $n_1, n_2 > 1$ and provides a total xWLN of 0.052 for $n_1, n_2\leq9$. Its output states and xWLN distribution are plotted at the bottom of Fig.~\ref{fig:GKPcompactVsHybrid}a and b. The choice of $\xi_0=\xi_1$, on the other hand, lowers the total xWLN to 0.034 highlighting the potential benefit of compromising between individual success rates and feed-forward compatibility. When a photon loss of $\eta=0.95$ is present, a choice of $\kappa=0.78$ (optimal for the underlying cat-like states) preserves a total xWLN of 0.025 ($\xi_0=0$), 0.030 ($\xi_0=\frac{1}{2}\xi_1$), and 0.020 ($\xi_0=\xi_1$) pointing to a comparable loss tolerance of all three setup variants. Following the discussion at the end of Sec.~\ref{sec:CatLikeStates}, the lossy \textit{hybrid} setups and thus the lossy \textit{compact} setup can be related to pure ones with an offline squeezing found to be 9.8 dB.
The feed-forward for both \textit{hybrid} setups was chosen based on the case $\xi_0=\xi_1$ and Eq.~\eqref{eq:MultiModeFF} resulting in
\begin{align}
    C_{\xi_0}\mapsto C_{\xi_0}\left(C^{-1}_{\xi_1}\cdot\left[\begin{smallmatrix}
        \sqrt{n_j/n_{j, \text{opt}}} &  \\
         & \sqrt{n_{j, \text{opt}}/n_j}
    \end{smallmatrix}\right]C_{\xi_1}\right),
\end{align}
where $n_{1, \text{opt}}=n_{2, \text{opt}}=4$, as well as
\begin{align}
    d_{\text{FF}, j}=\pm\tfrac{1-(-1)^{n}}{2}\left[
    \begin{smallmatrix}
        0 \\
        1.53/\sqrt{n_j}
    \end{smallmatrix}
    \right],
\end{align}
with signs selected to maximise the projection's success rates. Presumably, this can be optimised further for $\xi_0\neq\xi_1$.
Besides, the squeezing and displacement of the displayed output states were set to match those of the \textit{compact} setup.

\paragraph*{Composite setup.}\label{sec:CompositeSetup}
The breeding protocol can be reobtained starting from Eq.~\eqref{eq:BlurTotal}. Setting 
\begin{align}
    D_A^{(2)}=\text{diag}{\textstyle\left(\frac{\lambda^2}{\cosh\xi_1}, \frac{\lambda^{-2}}{\cosh\xi_1}, \frac{\lambda^2}{\cosh\xi_1}, \frac{\lambda^{-2}}{\cosh\xi_1}, ...\right)}
\end{align}
to commute with $O_A$ and
\begin{align}
    C^{(2)}=C^{(1)}\cdot \text{diag}\left(1, 1; s, s^{-1}, ...\right)
\end{align}
results in the two conditions
\begin{align}
    s^2\cdot \tanh^2\xi_1&\approx\cosh\xi_1 - \frac{1}{\cosh\xi_1},\\
    s^{-2}\cdot \tanh^2\xi_1&\approx0,
\end{align}
with the exact solutions $s=\sqrt{\cosh\xi_1}$ and $s=\infty$. Directly calculating $A^{(0)}$ reveals that the former underestimates $s$ and the \textit{composite} setup variant is indeed obtained for the high squeezing limit, which can be implemented by a post-selected homodyne detection. Figure~\ref{fig:GKPffComparison} shows the improvement in feed-forward performance when transitioning from \textit{compact} to \textit{composite} in the case of $n_1=2$ and $n_2=5$.

As high squeezing makes a vacuum projection unlikely to succeed, the \textit{composite} setup necessitates a Gaussian measurement with subsequent post-selection and feed-forward. This is achieved by combining the infinite squeezing and vacuum projection into a homodyne measurement in $p$. The resulting states for different measurement outcomes $p_0$ as well as different photon patterns $(n_1, n_2)$ are plotted in Fig.~\ref{fig:Homodyne_grid}.
Generally, we find that an increase in $p_0$ leads to a reduction in photons, i.e.\ fewer rows of negative peaks, with transitions between photon numbers causing non-grid states.
The latter occur roughly at even/odd multiples of the potential feed-forward displacement.
Furthermore, a higher variance of the photon pattern can be seen to increase the outputs' asymmetry.

\begin{figure}[t]
    \centering
    \includegraphics[width=\linewidth]{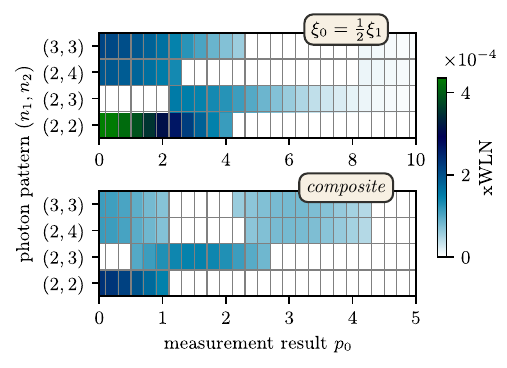}
    \caption{Distribution of xWLN of the \textit{hybrid} ($\xi_0=\frac{1}{2}\xi_1$) and \textit{composite} setups using homodyne detection for different measurement results $p_0$ and photon patterns $(n_1, n_2)$.}
    \label{fig:xWLN_Homodyne}
\end{figure}

\paragraph*{Composite vs hybrid setup.}
The \textit{hybrid} setup also benefits from replacing heterodyne with homodyne detection: Increasing $\xi_0$ raises the necessary offline squeezing, thus applying as much squeezing as possible via the Gaussian measurement helps minimise resources. For $\xi_0=\xi_1$, this leads to $D_A=D_A^{(2)}$, which commutes with $O_A$ and produces a setup which is mathematically equivalent to the \textit{composite} one. In the case of $\xi_0<\xi_1$, a $D_A$ closer to $D_A^{(1)}$ can be chosen while still retaining the maximum offline squeezing of 12.5 dB.
\textit{Hybrid} output states resulting from such a setup with $\xi_0=\frac{1}{2}\xi_1$ are displayed alongside the \textit{composite} ones in Fig.~\ref{fig:Homodyne_grid}. While both sets of outputs are similar, the feed-forward operations for $n_1\neq n_2$ are shown to introduce additional distortion for the \textit{hybrid} setup. At the same time, its measurement-induced transitions happen less frequently at about four times higher $p_0$.
After a naive post-selection discarding both transition as well as two-peaked states, the xWLN distribution generated by the two setups is plotted in Fig.~\ref{fig:xWLN_Homodyne}, while the total xWLN of the selected photon patterns is given in Table~\ref{tab:xWLN_homodyne}.
We find that setting $\xi_0=\frac{1}{2}\xi_1$ together with homodyne detection does indeed increase success rates whenever $n_1=n_2$, especially if the initial amount of negative peaks is already minimal. On the other hand, its xWLN is comparable to the roughly uniform level reached by the \textit{composite} setup when a feed-forward is involved, despite some visible deviation of the outputs. In the extreme, the \textit{compact} setup produces the individually highest success rates, but renders the output of most photon patterns unusable.
This trade-off is fully and continuously tunable by setting $0\leq\xi_0\leq\xi_1$ and can evidently be used to improve the performance of setups at either end of the spectrum.
These results emphasize that, even if an exact \textit{composite} setup variant exists, the combination of \textit{compact} and \textit{composite} properties provided by the \textit{hybrid} design can generally be beneficial.
Aside from $\xi_0$, the presented analysis highlights a small set of parameters relevant to the generation of GKP-like states which can be optimised over for any setup size $k$.

\begin{table}[t]
    \centering
    \caption{Post-selected xWLN for different photon patterns generated by setups using homodyne detection.}\label{tab:xWLN_homodyne}
    \begin{ruledtabular}
        \begin{tabular}{CCCC}
            (n_1, n_2) & \xi_0=0 & \xi_0=\frac{1}{2}\xi_1 & \xi_0=\xi_1 \vspace*{0.5mm}\\ \hline \rule{0mm}{\normalbaselineskip}
            (3, 3) & 0.0048 & 0.0033 & 0.0025 \\
            (2, 4) & 0.0042 & 0.0023 & 0.0024 \\
            (2, 3) & - & 0.0026 & 0.0026 \\
            (2, 2) & 0.0083 & 0.0060 & 0.0021
        \end{tabular}
    \end{ruledtabular}
\end{table}

\begin{figure}[t]
    \centering
    \includegraphics[width=\linewidth]{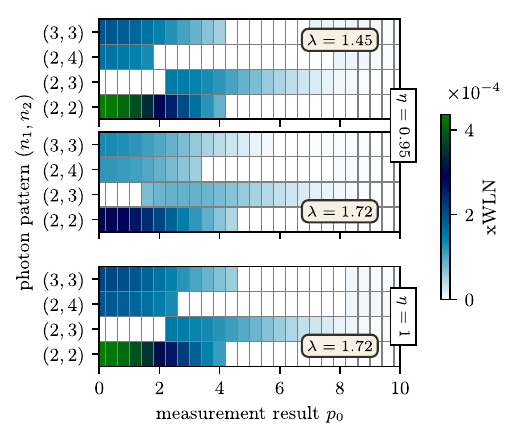}
    \caption{Performance of the loss-adjusted \textit{hybrid} setup with $\xi_0=\frac{1}{2}\xi_1$ and homodyne detection. Shown is the xWLN distribution for different measurement results $p_0$ and photon patterns $(n_1, n_2)$ in case of no loss ($\eta=1$, $\lambda=1.72$), loss acting on the unadjusted setup ($\eta=0.95$, $\lambda=1.72$), and loss acting on the adjusted setup ($\eta=0.95$, $\lambda=1.45$).}
    \label{fig:xWLN_Homodyne_loss}
\end{figure}

\paragraph*{Loss tolerance.}
Besides the two-mode squeezing $\vec{\xi}$ and the Gaussian measurements, photon loss also contributes to the blurring $A^{(0)}$. Consequently, it is sensible to adjust the former two to the given loss level. In the case of $\xi_0=\xi_1$, this simply corresponds to lowering $\lambda$ as demonstrated for the underlying cat-like states, otherwise the chosen $D_A$ also needs to be adapted. Here, we choose $\lambda=1.45$ which retains the medium blurring strength $(D_A)_{11}$ of the pure setup for a loss of $\eta=0.95$.
The resulting xWLN distribution for $\xi_0=\frac{1}{2}\xi_1$, a maximum offline squeezing of 12.5 dB and the different photon patterns is depicted in Fig.~\ref{fig:xWLN_Homodyne_loss} with and without this loss adjustment.
The corresponding total xWLN is given in Table~\ref{tab:xWLN_homodyne_loss} while the associated output states can be found in App.~\ref{sec:AppendixFigures}.
\begin{table}[h]
    \centering
    \caption{Post-selected xWLN for different photon patterns generated using $\xi_0=\frac{1}{2}\xi_1$ and homodyne detection while changing the amount of photon loss $\eta$ along with the asymmetry $\lambda$.}\label{tab:xWLN_homodyne_loss}
    \begin{ruledtabular}
        \begin{tabular}{CCCC}
            &\eta=1&\multicolumn{2}{C}{\eta=0.95}\\ \cmidrule{2-2}\cmidrule{3-4}
            (n_1, n_2) & \lambda=1.72 & \lambda=1.72 & \lambda=1.45 \\ \hline \rule{0mm}{\normalbaselineskip}
            (3, 3) & 0.0033 & 0.0022 & 0.0029 \\
            (2, 4) & 0.0023 & 0.0016 & 0.0014 \\
            (2, 3) & 0.0026 & 0.0019 & 0.0024 \\
            (2, 2) & 0.0060 & 0.0042 & 0.0058
        \end{tabular}
    \end{ruledtabular}
\end{table}
We find that the loss-adjusted setup significantly outperforms the unadjusted one for most outcomes and indeed comes close to the case of no loss. The only exception is the photon pattern $(2,4)$ for which a stricter post-selection results in a lower xWLN.
Note that $\kappa=\lambda^{-1}$ was left untouched as an optimisation over all viable outputs is tedious. Nevertheless, these results show that the adaptation of a few parameters -- all performed within the Gaussian formalism -- can already provide a strong starting point for further optimisations.
Furthermore, they demonstrate how a general understanding of different setup components as contributions to the total Gaussian blur $A^{(0)}$ can provide intuitive tools to adjust and optimise the generation of PSG states even for setups with large $k$.

%% file: conclusion.tex
\section{Conclusion}\label{sec:Conclusion}

We have introduced a mathematical formalism offering a Gaussian perspective on photon-subtracted Gaussian states
by writing their Wigner function as a Gaussian blur and filter acting on the Wigner function of the measured Fock states.
Within this formalism, the offline squeezing, Gaussian measurements, and photon loss of an experimental setup become distinct contributions to the total blur, enabling the efficient optimisation of general GBS-like setups by compromising between success rates, feed-forward compatibility, and loss tolerance.

For a single PNRD, we found that relevant states can be fully described by only two parameters. Analysing this two-dimensional state space then lead to improvements to the common cat as well as cubic phase state generation by adjusting the asymmetry $\lambda$ and displacement $\delta$.
Furthermore, we identified two distinct single-mode feed-forward schemes associated with the cat- and cubic phase-like output states and demonstrated how they can be used to obtain general multi-mode feed-forwards for setups with more than one PNRD.
For this purpose, a \textit{hybrid} setup variant was introduced which bridges the gap between \textit{compact} setups relying on post-selection and \textit{composite} breeding protocols and can be used to continuously morph between the two.
In the considered example of GKP state generation, this \textit{hybrid} version was shown to combine respective properties and consequently outperform the \textit{compact} as well as \textit{composite} setup variant, highlighting the benefits of compromising between individual success rates and feed-forward compatibility.

In the presence of photon loss, we showcased the importance of optimising the output squeezing $\kappa$ for setups with a single PNRD.
Moreover, we found that it is often possible to accurately replicate pure output states despite the presence of loss.
The necessary setup modifications are designed to keep the total blur $A^{(0)}$ constant and provide an effective tool for tailoring general setups to a given level of loss.
For multiple PNRDs, this was tested for GKP state generation where these loss-adjustments led to significant performance improvements.

Finally, we established a general lower bound on the resources needed to generate a given target state near-deterministically
based on the expected Wigner logarithmic negativity as an additive monotone of non-Gaussianity and natural figure of merit.

%% file: appendix.tex
\section{Derivation of Equation \eqref{eq:GaussianPSG}}
\label{sec:AppendixDerivation}

For a partitioned, positive definite matrix $M$, we find
\begin{align}
    &\begin{bmatrix}\vec{r}\\\vec{q}\end{bmatrix}^T\begin{bmatrix}M_A&M_{AB}\\M_{AB}^T&M_B\end{bmatrix}\begin{bmatrix}\vec{r}\\\vec{q}\end{bmatrix}=\left(\vec{r}+M_A^{-1}M_{AB}\vec{q}\right)^T M_A \left(\vec{r}+M_A^{-1}M_{AB}\vec{q}\right) + \vec{q}^T \left(M/M_A\right) \vec{q},
\end{align}
with the Schur complement $M/M_A = M_B - M_{AB}^TM_A^{-1}M_{AB}$. Setting $M=\Sigma^{-1}$, block-wise matrix inversion \cite{bernsteinMatrixMathematicsTheory2009} gives
\begin{align}
    M_A = \left(\Sigma/\Sigma_B\right)^{-1}&&\text{as well as}&& M_{AB}=-\left(\Sigma/\Sigma_B\right)^{-1}\Sigma_{AB}\Sigma_B^{-1}.
\end{align}
Conversely, it is $\Sigma_B^{-1}=M/M_A$ and we obtain
\begin{align}
    G_{\Sigma, \vec{\mu}}\left(\begin{bmatrix}\vec{r}\\ \vec{q}\end{bmatrix}\right)=G_{\Sigma, \vec{0}}\left(\begin{bmatrix}\vec{r}-\vec{\mu}_A\\ \vec{q}-\vec{\mu}_B\end{bmatrix}\right)
    =G_{\Sigma_B, \vec{\mu}_B}\left(\vec{q}\right) G_{\Sigma/\Sigma_B, \vec{\mu}_A - \Sigma_{AB}\Sigma_B^{-1}\vec{\mu}_B}\left(\vec{r}-\Sigma_{AB}\Sigma_B^{-1}\vec{q}\right)
    =G_{B, \vec{\beta}}\left(\vec{q}\right)G_{A, -\vec{\gamma}}\left(C\vec{q}-\vec{r}\right).
\end{align}
Equation~\eqref{eq:GaussianPSG} then follows directly from Eq.~\eqref{eq:GaussianPSG_0}.

\section{Figures}\label{sec:AppendixFigures}
\begin{figure}[htbp]
    \centering
    \includegraphics[width=\linewidth]{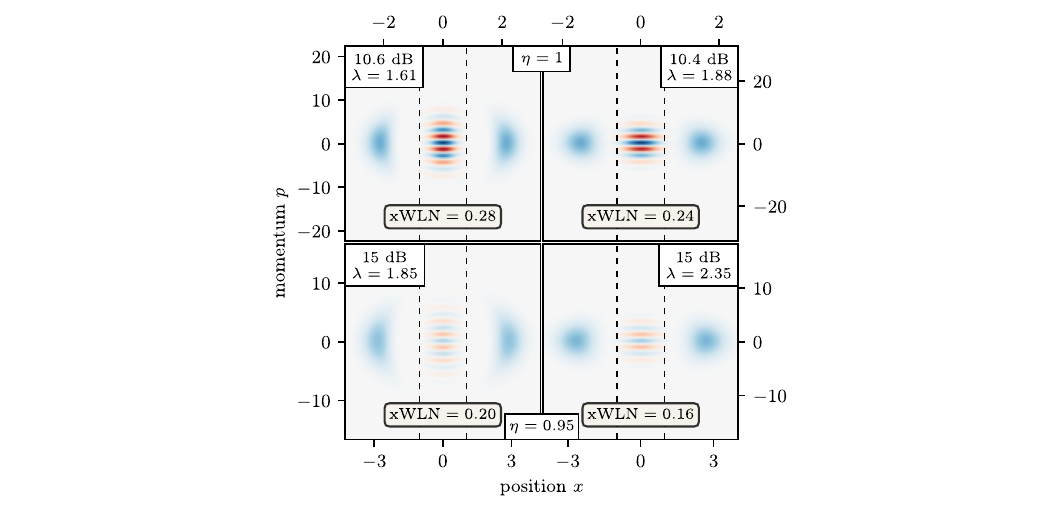}
    \caption{Comparison of two lossy cat-like states (bottom) with their related pure ones (top) for $n_1=8$. The photon loss of $\eta=0.95$ can be compensated for by reducing $\lambda$ as well as the used offline squeezing resulting in perfectly matched outlines of the negative regions of the respective Wigner functions.
    The marked region represents the two-sigma interval of the Gaussian blur and can be used to indicate a setup's feed-forward compatibility.}
\end{figure}

\begin{figure}[htbp]
    \centering
    \includegraphics[width=\linewidth]{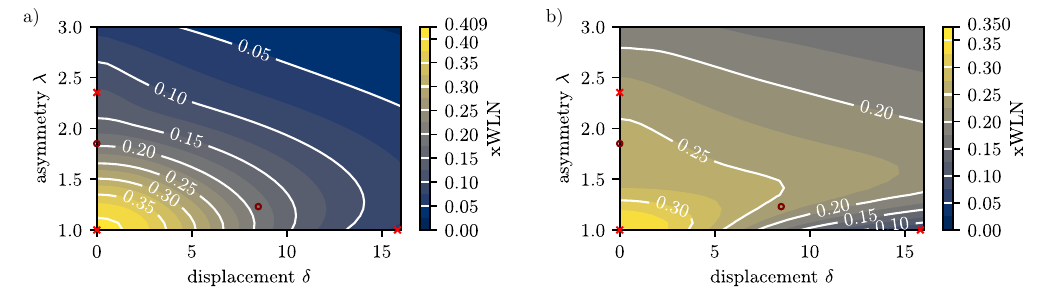}
    \caption{Expected Wigner logarithmic negativity (xWLN) for a GBS-like setup with a single PNRD, 15 dB offline squeezing, and an exclusive a) detector loss of $\eta_A=0.90$, b) output loss of $\eta_B=0.90$. The standard parameters for generating Fock-, cat-, and cubic phase-like states are marked with red crosses, while the custom choices shown in Figs.~\ref{fig:CatLike} and \ref{fig:CpsLike} are indicated by red circles.}
    \label{fig:Scan_B}
\end{figure}
\begin{figure}[htbp]
    \centering
    \includegraphics[width=\linewidth]{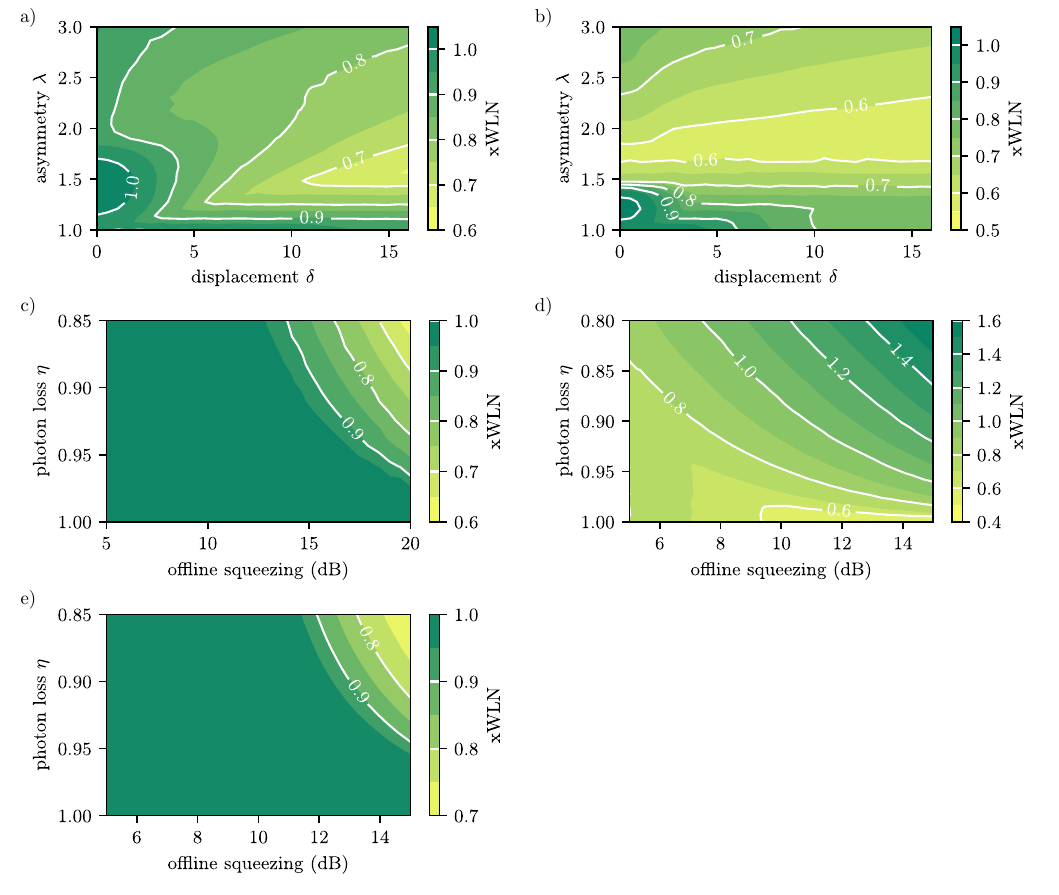}
    \caption{Optimal $\kappa$ values used for the xWLN calculations of a) Fig.~\ref{fig:Scan_95}, b) Fig.~\ref{fig:Scan_B}b, c) Fig.~\ref{fig:xWLN_Fock}, d) Fig.~\ref{fig:xWLN_Cat}, and e) Fig.~\ref{fig:xWLN_Cps}.}
    \label{fig:optB}
\end{figure}

\begin{figure}[htbp]
    \centering
    \includegraphics[width=\linewidth]{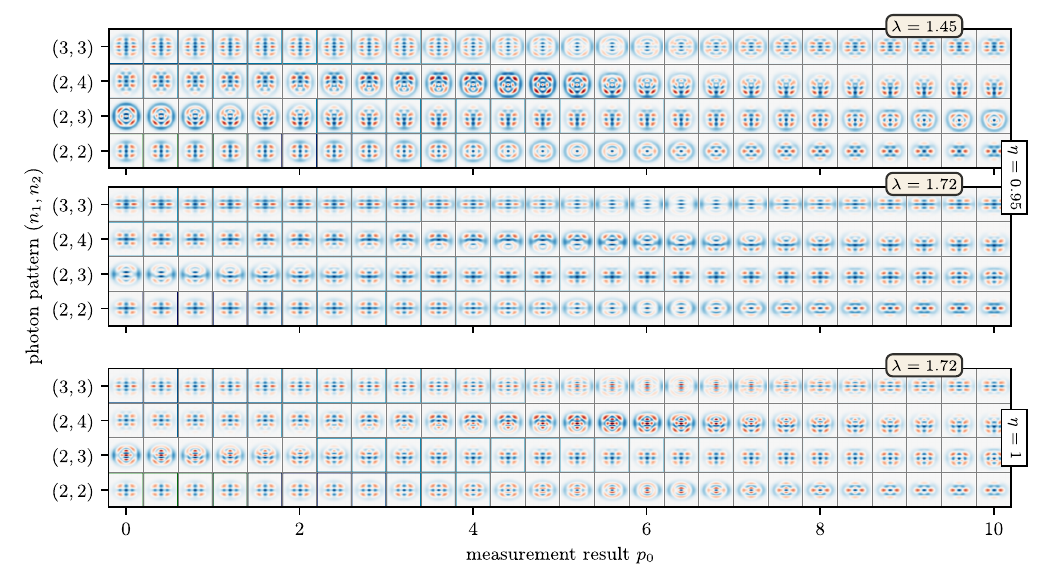}
    \caption{GKP-like output states of the \textit{hybrid} setup with $\xi_0=\frac{1}{2}\xi_1$ using homodyne detection for different measurement results $p_0$, photon patterns $(n_1, n_2)$, levels of photon loss $\eta$, and asymmetries $\lambda$.}
\end{figure}